# Deep Learning in Ultrasound Elastography Imaging

*Hongliang Li, Manish Bhatt, Zhen Qu, Shiming Zhang, Martin C. Hartel, Ali Khademhosseini and Guy Cloutier**

(* Corresponding authors: guy.cloutier@umontreal.ca (Guy Cloutier))

Dr. Hongliang Li, Dr. Manish Bhatt, Dr. Zhen Qu, Prof. Guy Cloutier
Laboratory of Biorheology and Medical Ultrasonics (LBUM), University of Montreal Hospital Research Center (CRCHUM), 900 Saint Denis, Montréal, H2X 0A9, Canada

Dr. Shiming Zhang, Martin C. Hartel, Prof. Ali Khademhosseini
California Nanosystems Institute, University of California, Los Angeles, 570 Westwood Plaza, Los Angeles, CA 90095, USA

Keywords: (ultrasound, elastography, artificial intelligence, deep learning)

**Abstract:** It is known that changes in the mechanical properties of tissues are associated with the onset and progression of certain diseases. Ultrasound elastography is a technique to characterize tissue stiffness using ultrasound imaging either by measuring tissue strain using quasi-static elastography or natural organ pulsation elastography, or by tracing a propagated shear wave induced by a source or a natural vibration using dynamic elastography. In recent years, deep learning has begun to emerge in ultrasound elastography research. In this review, several common deep learning frameworks in the computer vision community, such as multilayer perceptron, convolutional neural network, and recurrent neural network are described. Then, recent advances in ultrasound elastography using such deep learning techniques are revisited in terms of algorithm development and clinical diagnosis. Finally, the current challenges and future developments of deep learning in ultrasound elastography are prospected.

## 1. Introduction

Changes in the mechanical properties of tissues are associated with the onset and progression of some diseases. For example, liver fibrosis is associated with an increased liver tissue stiffness, and manual palpation has been used for centuries by physicians to compress an organ and feel its stiffness for disease diagnosis. However, interpreting palpation remains subjective to physicians. Ultrasound elastography seeks to objectively characterize tissue stiffness using ultrasound imaging. Specifically, tissue deformation is induced by internal or external vibrations. Then, the response of the tissue deformation is captured by ultrasound images from which elastography techniques can derive information on tissue stiffness.

Ultrasound techniques for mechanical tissue assessment have been very actively studied over the years and can be divided into two groups: quasi-static and dynamic elastography. The quasi-static methods can determine the elasticity or Young's modulus $E$ by measuring the strain $\varepsilon$ caused by a known compressive force $\sigma$ and the area of such force. Since it is challenging to measure the compressive force locally at the site of deformation, scientists usually only measure the strain to provide contrast in terms of relative stiffness. As an alternative in quasi-static elastography, the natural pulsation of an organ is often used and strain $\varepsilon$ is measured to assess its rigidity. Dynamic methods in elastography can determine the complex shear modulus from which the elasticity $E$ and viscosity $\eta$ can be derived by tracing and analyzing a propagated shear wave into the specimen. Since the shear modulus of various tissues spans over orders of magnitude, dynamic methods provide high contrast among different tissues/organs regarding their mechanical properties. The source of vibration in shear wave elastography (SWE) can be external (actuator), internal (radiation pressure) or natural (e.g., shear waves produced by closing heart valves).

In recent years, deep learning has been widely developed by the computer vision community. However, the application of deep learning to analyze medical images is not straightforward. One reason is that the labeled medical data for training is not as readily available as computer vision images due to the higher costs and professional workload needed for their procurement. Ethical regulatory issues for use of medical images also have to be considered. In addition, medical images are more complex and less interpretable than scenic images. In the case of ultrasound, the inter- and intra-observer variabilities limit the clinical diagnosis capabilities using deep learning. Even so, more and more efforts have been explored in ultrasound elastography research. The objective of this review is to focus on recent advances in ultrasound elastography using deep learning techniques in terms of algorithm development and clinical diagnosis.

The rest of the paper is divided into four sections: (i) an overview of ultrasound elastography, (ii) an overview of deep learning techniques used in ultrasound elastography, (iii) a literature review on the applications of deep learning in ultrasound elastography, and (iv) a summary and perspectives.

## 2. Ultrasound elastography

Most soft bio-tissues can be considered as fluid-like solids, as they contain more than 70-w% of water, and thus exhibit mechanical characteristics of both solids and fluids [1]. One way to characterize them is to analyze their viscoelastic properties.

Elasticity is a physical property that describes the ability of a material to return to its original shape after a stress is removed [2], and it mainly reflects the solid properties of the material. Three major moduli are used to describe the elasticity of a material: the bulk modulus $K$, Young's modulus $E$, and shear modulus $G$. The modulus $K$ is defined as the ratio of the pressure and volume changes. The Young's modulus $E$ corresponds to the ratio between tensile stress $\sigma$ and tensile strain $\varepsilon$ of a material, $E = \sigma / \varepsilon$ [3], while $G$ defines the ratio of shear stress and shear strain of a material. For an incompressible material (or a purely elastic material, which applies to many soft tissues as they are nearly incompressible), it can be assumed that $E = 3G'$, where $G'$ is the real component of $G$ (while $G = G'$ for a purely elastic material), also known as the shear storage modulus [4, 5]. Besides, $E$ and $G$ also have a broader dynamic range over different types of human tissues than $K$ [1]. The comparison between $K$ and $E$ (or $G$) is shown in **Figure 1**. Thus, $E$ and $G$ are more discriminant and of more interest/significance for the characterization of human soft tissues.

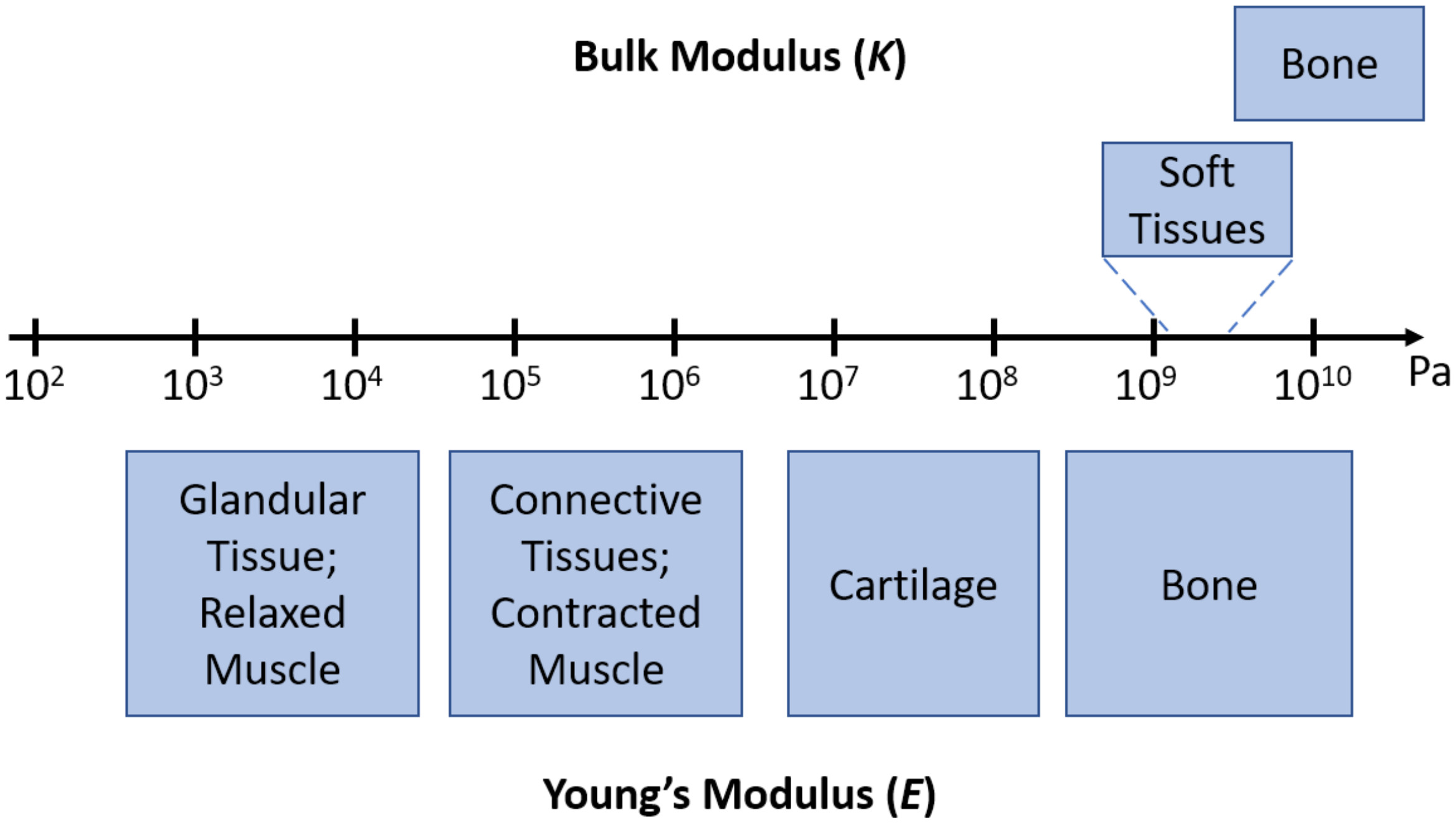

**Figure 1.** Moduli variance with different types of human tissues.

On the other hand, the viscosity $\eta$ describes the ability of a material to resist to the deformation due to tensile stress or shear stress [2], and it mainly reflects the fluid-like properties. The shear lost modulus ($G''$, which represents the imaginary component of $G$ can be used to describe the viscous behavior of a tissue). Alternatively, a Voigt mechanical model, as shown in **Figure 2**, is frequently used to describe viscoelastic materials. It is represented by a purely elastic spring connected in parallel to a purely viscous damper. Such a model can be used to describe either a tensile viscoelastic material or a shear viscoelastic material. A material with only elasticity is called a purely elastic material, while a material with only viscosity is called a Newtonian fluid. A soft tissue or soft tissue-like material falls between these two extreme conditions and can be called a viscoelastic material.

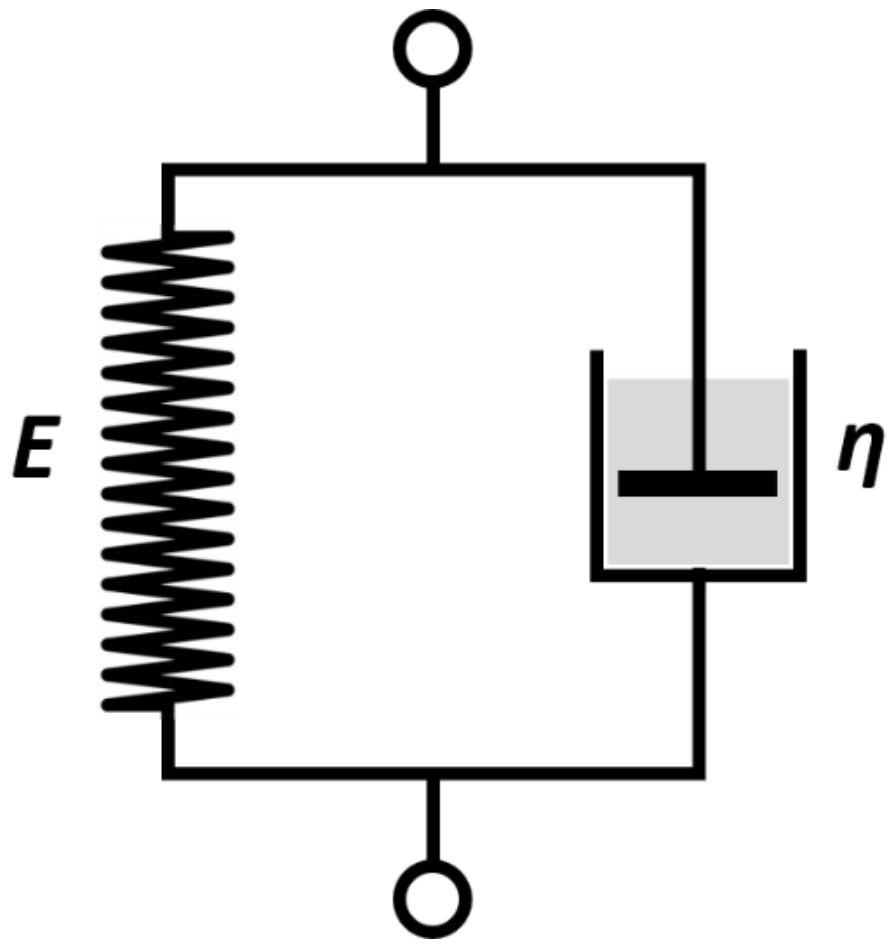

**Figure 2.** The Voigt material model described by a spring with a Young's modulus *E* and a dashpot in parallel with a viscosity $\eta$.

The popular methods for determining a material's elasticity or viscosity are all based on the same objective, to detect the response of the target tissue after an excitation. Once the response is detected, the viscoelastic parameters can be derived from the mathematical relations between the response and the excitation [6]. Those methods could be divided into two groups according to the temporal difference of the excitation: quasi-static measurements and dynamic measurements (also termed shear wave measurements).

### 2.1. Quasi-static elastography

Quasi-static elastography, also called strain imaging, measures the tissue strain under a stress induced by a compression and/or a relaxation. The first work on strain imaging utilized a manual compression on the top of a tissue to induce its deformation [7]. The schematic and concept of strain imaging is shown in **Figure 3**. Here, an ultrasound transducer uniaxially applies a small compression on a soft phantom with a hard inclusion. The deformation process of the phantom can be regarded as the compression of a spring, where the parts with different stiffnesses give rise to different displacements according to Hooke's law. Since the stress $\sigma$ may not be measured in practice, the absolute value of *E* may not be determined. Instead, a relative stiffness is determinable, i.e. higher $\varepsilon$ represents softer whereas lower $\varepsilon$ represents stiffer materials. To measure these displacements at each depth, a cross-correlation method is usually applied to analyze two radio-frequency (RF) echoes, namely pre-compression and post-compression signals. The position which provides the largest value in a cross-correlation map determines the displacement between pre-compression and post-compression signals. Finally, the spatial gradient of the displacement field represents the strain distribution, which is also known as an elastogram.

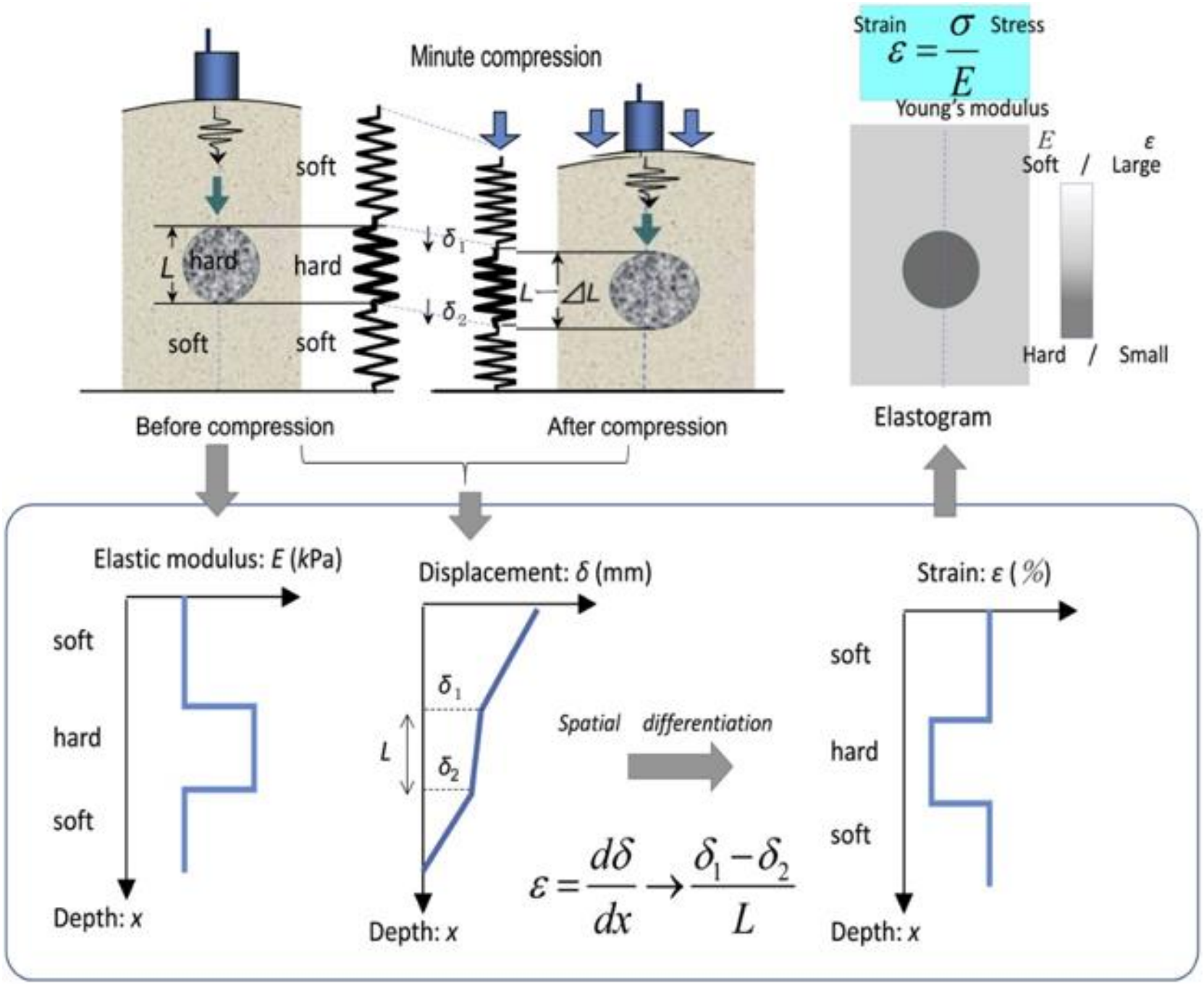

**Figure 3.** Principle of ultrasound strain imaging. A minute compression is applied on a soft phantom with a hard inclusion. Two RF echoes, described as pre-compression and post-compression signals, are acquired and analyzed to obtain displacements at each depth. Finally, the spatial gradient of the displacement field represents the strain distribution. Reproduced with permission from [8]. Copyright 2015, Elsevier.

Another way to induce deformation is to exploit natural cardiovascular or respiratory movements. Likewise, previous displacement estimation methods such as the cross-correlation are also applicable to strain imaging [9-11]. However, the most commonly used cross-correlation method only provides displacements. An additional gradient operation on displacements is needed to derive strains, which could induce estimation variances due to signal decorrelation in the presence of noise. Utilizing an affine model-based method is an alternative way to estimate strains directly. In the framework of the Lucas-Kanade optical flow estimation method, a Lagrangian speckle model estimator (LSME) was proposed to directly obtain affine motion components including displacements, strains and shears [12, 13]. Recently, the same group proposed a sparse model strain estimator (SMSE) to estimate a dense strain field at a high resolution and a high computation efficiency [14].

### 2.2. Dynamic elastography

Dynamic elastography uses shear wave (SW) measurement methods to determine the tissue viscoelasticity by tracing and analyzing the propagating wave into the specimen. Shear waves, also called transverse waves, are moving mechanical waves that consist of particle oscillations occurring perpendicular to the direction of the energy transfer. In a viscoelastic material, the SW propagation equations are given by [15]:

$$\rho\frac{\partial^2 u_i}{\partial t^2} - \mu\nabla^2 u_i = 0, \quad (1)$$

$$\mu = G' + jG'', \quad (2)$$

where $\rho$ is the density of the specimen, $u_i$ is the particle displacement occurring in the $i$ direction, and $t$ is time. $\nabla^2 u_i$ is the spatial Laplacian of the particle displacement. The parameter $\mu$ is the second Lamé coefficient and it denotes the shear viscoelasticity represented by the shear storage and loss moduli, $G'$ and $G''$, respectively, in which $G'$ reflects the elastic property and $G''$ the viscous response.

Based on the Voigt model, $G'$ and $\eta$ have the following relationship with the velocity $v_S$ and the attenuation $\alpha_S$ of the SW, where $\omega_S$ represents the angular frequency of the SW [16]:

$$G' = \frac{\rho\omega_S{}^2 v_S{}^2 \cdot (\omega_S{}^2 - \alpha_S{}^2 v_S{}^2)}{(\omega_S{}^2 + \alpha_S{}^2 v_S{}^2)^2}, \quad (3)$$

$$\eta = \frac{\rho\omega_S{}^2 v_S{}^2 \cdot 2\alpha_S v_S}{(\omega_S{}^2 + \alpha_S{}^2 v_S{}^2)^2}. \quad (4)$$

It is worth noting that for a soft biological tissue, the elastic property dominates the tissue mechanical property and the viscous behavior is often neglected. Therefore, in such case, it is considered as a purely elastic material and $\alpha_S$ could be ignored, so that a specific material model is not necessarily required and thus $G'$ can be expressed by:

$$G' = \rho v_S{}^2. \quad (5)$$

When the viscosity $\eta$ is taken into account for the tissue characterization, according to **Equation (4)**, it can be determined either using both $v_S$ and $\alpha_S$, or alternatively by evaluating the dispersion of $v_S$ with respect to $\omega_S$ without determining the value of $\alpha_S$, i.e., solving **Equation (3)** by knowing multiple pairs of $v_S$ and $\omega_S$ [17]. Alternatively, if no rheological model is assumed (e.g., the Voigt model), then it may be more convenient to assess the shear storage $G'$ and loss $G''$ moduli to describe the mechanical property of the tissue [18]. This is particularly relevent considering the fact that the simple Voigt model might not allow mimicking both normal and pathological tissues of different grades of severity.

### 2.3. Technical review of elastography methods

In the past decades, many notable methods and techniques have been proposed and developed to quantify and visualize the tissue viscoelasticity using ultrasound [4, 19-21]. They typically use acoustic transducers to collect the information of the compressive force or SW excitation response by means of the tissue displacement. In such a technique, the field of view (FOV) is limited by the penetration depth of the ultrasound beam. Thus, if the region of interest (ROI) is in a deeper area, occluded by air, or attenuated by adipose tissues in the case of obese patients, it may not be reached [22].

#### *2.3.1. Quasi-static elastography developments*

Jonathan Ophir and his colleagues in 1991 were among the firsts to propose and describe this quasi-static measurement method [7]. Considering that the method indirectly finds the elastic property of the tissue, it was named as "elastography" at that time. It is worth noting that after a widely referred review paper by Gao et al. in 1996, the term "elastography" was expanded to describe all types of elasticity imaging methods [23]. It should also be noted that with a quasi-static measurement, tissue viscosity can also be determined by knowing the elasticity and measuring the relaxation time of the tissue from the compressive force [24-26]. Later on, strains of the cardiac muscle were measured with natural cardiac pulsation as the mechanical stimulus, which is referred to myocardial elastography [27]. Elastography was also explored to characterize the vessel wall with intravascular ultrasound (IVUS) images. The natural cardiac pulsation or the inflation of a compliant balloon on a catheter was used to differentiate the coronary calcified plaque component (less strain) from non-calcified tissues (higher strain) *in vivo* [28, 29]. In 2004, non-invasive vascular elastography (NIVE) was proposed to show the potential of determining the stability or vulnerability of an atherosclerotic plaque using ultrasound imaging [12]. NIVE could determine the strain field of the moving vessel wall of a superficial artery (e.g., the carotid) caused by the natural cardiac pulsation without the need of an external compression. This method was also introduced to study normal vessel walls [30]. Due to the limited lateral resolution of ultrasound images compared with the axial resolution, the motion in the lateral direction is challenging to track. Angular compounding [9], coherent plane wave compounding imaging [13], and synthetic aperture imaging [31] were proposed to enable more accurate lateral estimations, respectively. Out-of-plane motion is another potential factor causing estimation artifacts in the context of NIVE [32]. In addition, it is not easy to perform 2D estimations in the frequency domain as lateral phase information is not available since there is no carrier frequency in the lateral direction with conventional ultrasound images. As an alternative, a dedicated transverse oscillation image reconstruction method [33, 34] can be used for image reconstruction before strain estimations, as in [35].

*2.3.2. Dynamic elastography developments*

The dynamic methods derive the viscoelastic parameters according to **Equations (1)-(5)**, by tracing and analyzing the SW propagation in a specimen of interest. In 1988, Lerner et al. proposed a method to map the propagation of a low frequency continuous SW with a Doppler ultrasound displacement detection technique [36], and such an image of the SW displacement could reflect the tissue stiffness. Later in 1990, Yamakoshi et al. proposed a measurement method using an external mechanical vibration source, and $v_S$ was determined by analyzing the wavelength of a continuously propagated SW through an ultrasound Doppler technique [37]. Instead of measuring continuous SWs, Catheline et al. developed in 1999 an impulse SW measurement method, which used a high frame-rate ultrasound imaging system to trace the wavefront of the transient SW [38]. With this method, an ultrasonic probe is located at one side of the specimen to capture the propagation of the impulse SW that is generated by a mechanical vibrator located at the other side of the specimen. Conventional focused ultrasound systems are triggering transducer elements sequentially from one side of the image to the other side producing a frame-rate of typically below 100 frames/sec. The development of plane wave beamforming allowed implementing SW imaging methods more efficiently. Indeed, a plane wave system triggers all transducer elements of the probe at the same time to emit a plane wavefield, enabling it to have a high frame-rate (more than 1,000 frames/sec and up to 10,000 frames/sec) [39]. Using this approach, $v_S$ is determined through a time-of-flight technique by measuring the time elapse of the impulse SW to travel through two locations with a known distance. Since such a method used impulse SW, it is also called transient SW imaging, or transient elastography.

The SW excitation can also be made inside the specimen by means of an ultrasound radiation force (or acoustic radiation force, ARF) technique [15, 19, 21, 22, 40-42]. Among all, the supersonic shear imaging (SSI) technique [39] is considered the most advanced SW measurement method. With this method, the ultrafast plane wave imaging method is used to track multiple ARF impulses that are excited consecutively and very quickly at different depths. Each excitation produces a point source-like SW, which then interfere constructively and result in two SW planes propagating in opposite directions, as can be seen in **Figure 4**. By using the ultrafast scanner to trace these SWs, a two dimensional (2D) shear wave speed image of $v_S$ can be obtained. Moreover, since this technique creates broadband SWs, tissue viscosity can also be estimated through the $v_S$ dispersion method [22].

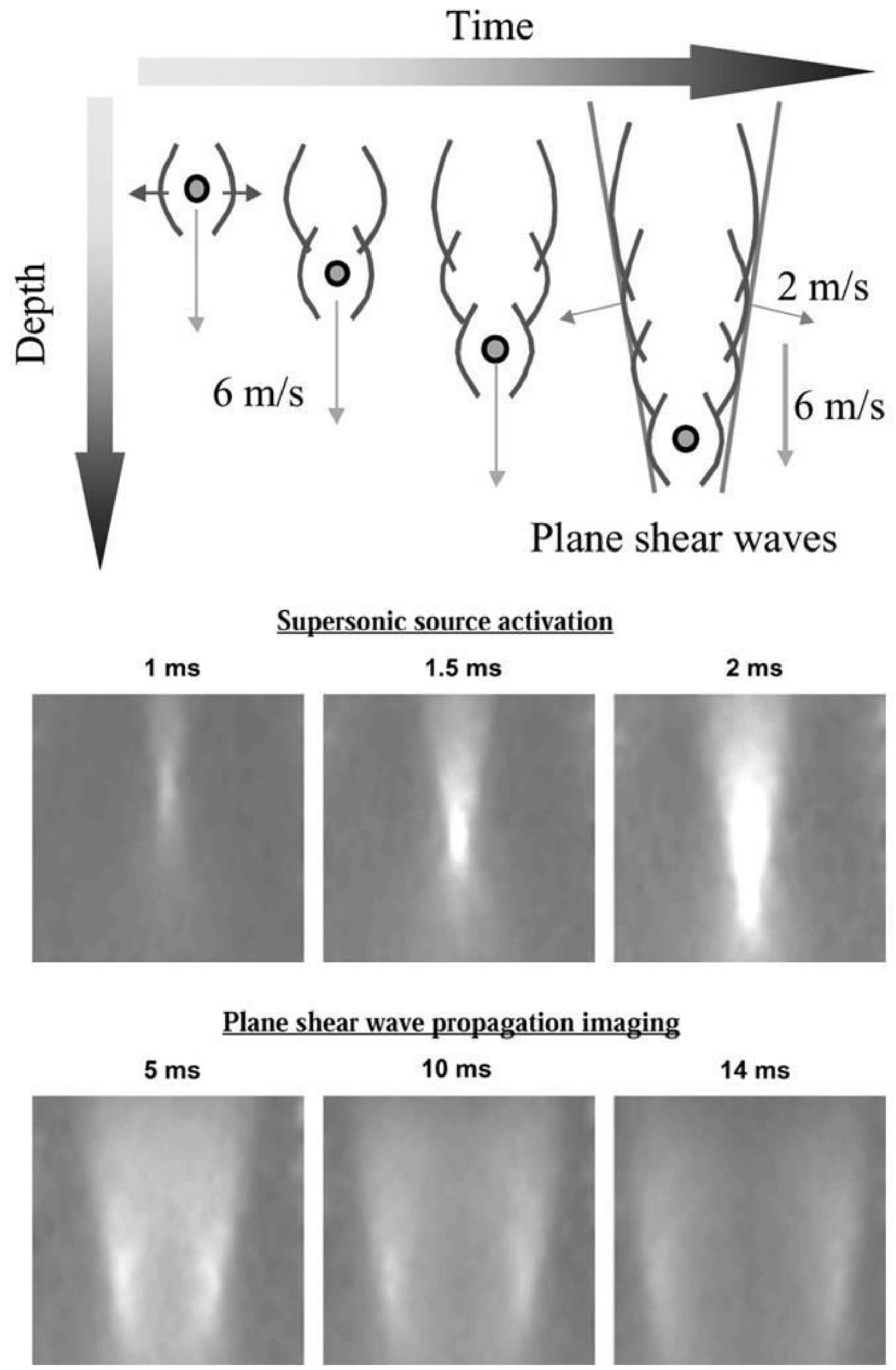

**Figure 4.** Generation of supersonic plane shear waves. Reproduced with permission from [39]. Copyright 2004, IEEE.

Over the past several years, many notable methods have been validated with different clinical datasets to diagnose several organ dysfunctions. For example, SW elastography was used to evaluate liver stiffness and diagnose patients with liver fibrosis and steatosis [42, 43]. Breast SW elastography could provide quantitative or qualitative information on breast lesions, and was widely applied to diagnose breast cancer [44, 45]. Dynamic elastography has also been shown as a good complementary method to detect the potential lesion area in prostate cancer [46]. When the lesion is identified, elastography may help clinicians make decisions on whether to carry out a biopsy in the lesion area. Additionally, dynamic elastography has been employed to differentiate thyroid cancers by evaluating the tissue stiffness [47]. Other applications, such as the characterization of the musculoskeletal system [48], Achilles tendon [49], lymph nodes [50], and many other organs [51-54], have also shown good results that reflect differences between normal and abnormal tissues.

At present, improving the measurement precision and accuracy of existing techniques is always an interesting and active topic. In addition, the most discussed parameter for clinical applications is still tissue elasticity, while viscosity is considered relatively less. Therefore, the development of a comprehensive measurement system for human soft tissue characterization including viscosity, anisotropy, and nonlinearity are interesting topics of elastography imaging [55-58], yet the innovations in the field have not received clinical approval.

## 3. Deep learning technologies

Deep learning is a special kind of artificial neural network (ANN) that tries to resemble the multilayered human cognition system. Even though ANN was first introduced in 1950, the development of deep learning methods took time primarily due to the lack of computing power and relevant data to train the systems. However, many limitations have been resolved and recent approaches have demonstrated comparable results to human performance in some situations. For example, image analysis of structures and abnormalities, tumor classifications, disease categorizations, and even composing preliminary radiology reports can now be handled well by deep learning tools. Previous computer-aided detection systems that were developed in the early 2000s for clinical applications were found to generate more false positives than humans [59]. It is expected that deep learning technology may help overcome the limitations of such systems by providing better detection accuracy.

The majority of the algorithms since 2015 employ supervised learning methods, namely convolutional neural networks (CNN) [60]. The advancements in hardware development and the availability of large labeled datasets (i.e., datasets with known indentification) resulted in improved CNN performance, encouraging their widespread use in medical imaging. The advantage of a deep neural network is to automatically learn significantly low level features such as lines or edges, and combine them with higher level features such as shapes in the subsequent layers. Krizhevsky et al. [61] introduced several concepts such as rectified linear unit (RELU) functions in CNNs, data augmentation, and dropouts, which have become dominant architectures used in current medical imaging strategies. CNNs provide advantages by requiring fewer trainable parameters, fewer training examples, and a lesser training time. Another popular type of network is the recurrent neural network (RNN). RNNs contain an internal memory state that can store information about previous data points and are often used in sequential data such as text or speech [62, 63]. Generator adversial networks (GANs) are another class of deep learning architecture that contain two networks: a generator and a discriminator [64, 65]. The generator and discriminator networks are trained jointly with a backpropagation algorithm. The training makes the generative network better at generating more realistic samples and the discriminator network better at differentiating artificially generated samples. GANs have found applications for image reconstruction in medical imaging.

### 3.1. Multilayer perceptron

The multilayer perceptron (MLP) is a deep feedforward neural network. A MLP network includes three components, an input layer, one or more hidden layers, and an output layer. The neural nodes from adjacent layers have connections, which allow information to be transferred from the input layer to the output layer without feedback. Other than connections, each neural node is associated with a weight and an activation function. The weights are learned during a training process. The activation function determines whether the output of a neural node should be activated, and the functional property can be linear or nonlinear.

### 3.2 Convolutional neural networks

Convolutional neural networks (CNNs) are the most successful and popular deep learning architecture in medical imaging research due to their ability to preserve spatial relationships when filtering input images. Spatial relationships in medical imaging define tissue interfaces, structural boundaries, or joints between muscle and bone; and thus, CNNs have found special attention. A CNN may take an input image of raw pixels and transform it via convolutional layers, rectified linear layers, and pooling layers. This is then fed into a fully connected layer, which assigns probabilities or class scores to categorize the input into the class with the highest probability. Pixels in an image can be processed by a different weight in a fully connected layer, or alternatively, every location can be processed using the same set of weights to extract various repeating patterns across the entire image. These trainable weights are referred to as kernels or filters; and are applied using a dot product or convolution and then processed by a nonlinearity function. Each convolutional layer may have several filters to extract multiple sets of patterns at each layer. Each convolutional layer is often followed by a pooling layer that reduces dimensionality and imposes translational invariance. These convolutional and pooling layers can be stacked together to form a multilayer network that often ends in one or more fully connected layers. A schematic diagram of a convolutional neural network is shown in **Figure 5**. The components of a CNN are briefly discussed below.

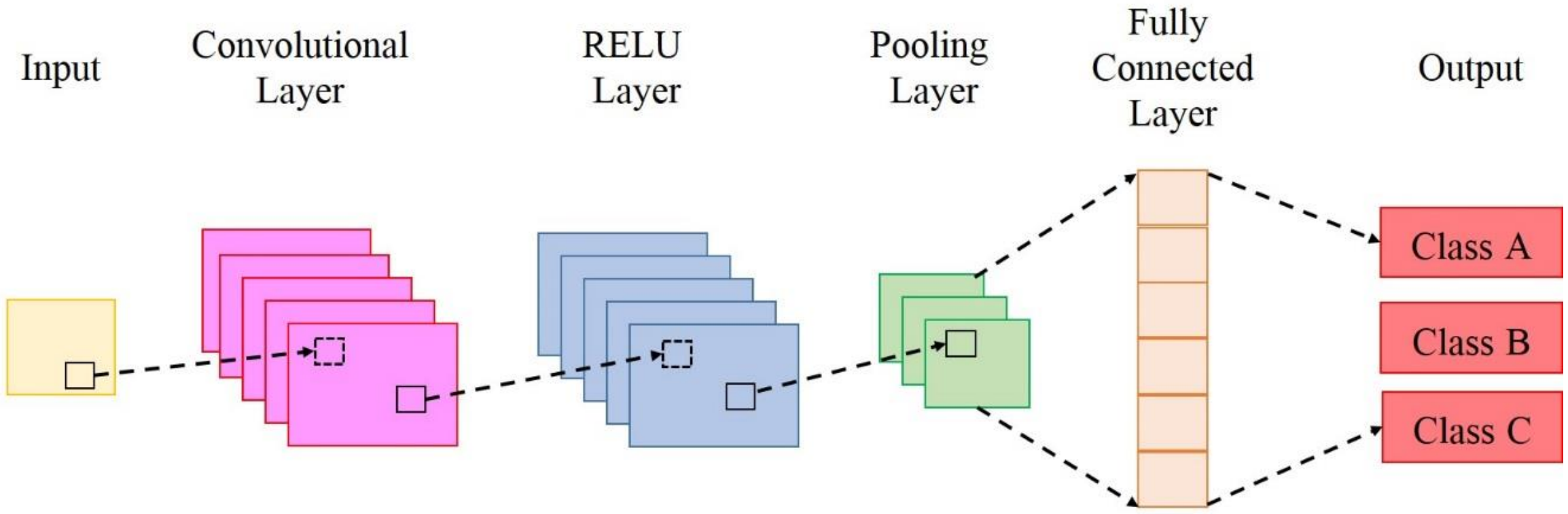

**Figure 5**. A schematic diagram of a convolutional neural network that has one convolutional layer.

*3.2.1. Convolutional layer*

The convolution layer performs a convolution operation between the function that consists of input values $I(t)$ (e.g., pixel values of an image) and another function that is a filter (or kernel) $K(t)$. Thus, the output $s(t)$ can be defined as $s(t) = I(t) * K(t)$. Convolution operations promote computationally efficient machine learning mainly due to their (a) sparse connections, (b) parameters (or weights) sharing, and (c) equivariant (or invariant) representation [66]. The CNN neurons make sparse connections, which means only some inputs are connected with the next layer. Convolutional neural networks reduce memory requirements by using each filter with its fixed weights across different positions of the input image. This is known as parameter sharing. Parameter sharing can improve the quality of the equivariant representation, meaning that input translations result in a corresponding feature translation. For a 2D image, a 2D convolution operation with input $I(m,n)$ and a kernel $K(a,b)$ can be defined as

$$s(t) = \sum_a \sum_b I(m-a, n-b) \,.\, K(a,b), \quad (6)$$

where $m, n$ denote the 2D image size and $a, b$ present the kernal size.

*3.2.2. Rectified linear unit layer*

The function of a rectified linear unit (RELU) layer is to set negative input values to zero to simplify and accelerate network computations and training. Mathematically, it can be expressed as

$$f(x) = \max(0, x). \quad (7)$$

where $x$ is the input. Other relevant activation functions include sigmoid, tanh, leaky RELUs, randomized RELUs, and parametric RELUs.

*3.2.3. Pooling layer*

The pooling layer is inserted in the network to reduce the dimensionality as well as the number of parameters to be computed. A common pooling method is max pooling, in which the largest input value within a filter is accepted and other values are discarded. This effectively summarizes the strongest activations over a neighborhood. The relative location of a strongly activated feature is considered more relevant than its exact location. Other pooling methods include average pooling and L2-normalization pooling.

#### *3.2.4. Fully connected layer*

The final layer of a CNN is the fully connected layer, where every neuron in the preceding layer is connected to every neuron in the fully connected layer. There can be more than one fully connected layer in a CNN depending on the level of feature abstraction desired. The fully connected layers take input from the preceding layer to compute the probability score by looking at the most strongly activated features and perform classification accordingly.

### 3.3 Recurrent neural networks

Recurrent neural networks, generally considered a type of supervised deep network, are mainly utilized for sequential data analysis, such as in text or speech processing or data with temporal information. The depth of an RNN can be as long as the input sequence length. The output of a layer in a plain recurrent neural network is connected to the next input as well as fed back into the layer itself, resulting in a capacity for contextual 'memory'. One of the major applications of RNNs is in image segmentation [67].

The structural characteristics of the RNN give it an inherent advantage for modeling sequence data [68, 69]. RNNs were not widely utilized until recently due to difficulties in training them to capture long term dependencies. The new modifications of RNNs have evolved into the long short term memory (LSTM) network and gated recurrent units (GRUs). These modifications allow holding long term dependencies or discarding some of the accumulated information.

### 3.4 Common problems and solution strategies in deep networks training

One of the major factors which determines success in deep learning performance is the availability of a large labeled dataset. However, in the field of medical imaging in general and in ultrasound elastography in particular, this requirement has been difficult to meet. Several reasons include the requirement of a clinical expert in this specialized field, associated high costs, ethical, legal and governmental legislation issues, and having a small dataset for some diseases that are not so common. Most medical imaging studies struggle to recruit more than 1000 patients for data collection, while deep learning applications in computer vision or robotics may include well beyond 1 million labeled images. Therefore, having to train a deep neural network with limited training data is a challenge. A very common issue that arises due to training with limited data is overfitting. To avoid overfitting, several strategies have been developed that include well-designed initialization [70], stochastic gradient descent optimization [71], efficient activation functions [72, 73], dropouts [74], and other powerful intermediate regularization methods [75].

In addition, small datasets also suffer from the class imbalance problem, i.e. the number of samples in one class being not comparable to another class. It is known that an imbalance in the training set may reduce the accuracy of the ANN. Practically, this means that when collecting medical images, the number of normal/healthy image results may usually be significantly more than those with abnormal/unhealthy results. Many techniques are available to deal with this challenge, one can be to undersample (i.e. reduce the amount of the majority class group), to oversample (i.e. duplicate some samples of the minority class), or combine these two approaches. Data augmentation is also an effective weapon to combat the problem: one can artificially generate more samples without actually collecting new data. The most commonly used methods include cropping/padding, blurring, and rotating parts of the existing images. Although image transformation is a common approach for data augmentation, an under covered challenge in ultrasound imaging and especially ultrasound elastography is to consider the physics of wave propagation to better represent artifacts in image formation (e.g., wave attenuation, wave reverberation, wave diffraction, etc … impacting the quality of clinical images), and non-rigid motions. Alternatively, one may also attempt to assign different weights to the cost function of different classes, so as to compensate for the imbalance issue, i.e. make it more costly when a minority sample is misclassified.

Another key solution is transfer learning, in which training of the network is performed with a partially related or unrelated dataset along with a labeled training dataset. The weights learned during the training of a CNN with the partially related or unrelated dataset are transferred to another CNN, which can be trained with the labeled medical dataset using these weights. Transfer learning can be divided into three categories: (a) Inductive transfer learning where regardless of whether target and source domains are the same or not, the target and source tasks are different; (b) Transductive transfer learning, where target and source domains are different while the target task is the same as the source task; and (c) Unsupervised transfer learning, where target and source domains may not be the same while the target task differs but is related to the source task [76]. Tajbakhsh et al. [77] proposed pre-training a CNN on a dataset of labeled natural images and then fine tune it using medical images. The study reported that a layer-wise fine-tuning scheme could improve the performance of a CNN network for the application at hand based on the amount of available data. Ravishankar et al. [78] observed that the detection performance depended upon the extent of the transfer and their study reported a 20% higher performance with a fine-tuned CNN. One of the main advantages of transfer learning is that it reduces expensive data-labeling efforts.

## 4. Deep learning applications in ultrasound elastography

### 4.1. Quasi-static elastography applications

In the last five years, deep learning has encouraged the development of new quasi-static elastography algorithms. For motion estimation problems, deep learning was originally applied to optical flow (OF) estimations in computer vision, in which a CNN-based architecture was constructed to perform end-to-end optical flow estimation without feature extractions, as shown in **Figure 6**. The first network called FlowNet [79] was proposed in 2015, which demonstrated competitive accuracy with conventional state-of-the-art OF algorithms, but with a higher computation efficiency. Following it, other CNN-based networks were developed, such as FlowNet 2.0 [80], SpyNet [81], LiteFlowNet [82] and PWC-Net [83], to achieve better accuracy. To our knowledge, the first two contributions which utilized a CNN-based OF architecture for ultrasound quasi-static elastography were published in 2018. One used a retrained FlowNet 2.0 with an additional ultrasound dataset to estimate breast strains [84]. The other study utilized FlowNet 2.0 directly to get a coarse estimate, then the initial displacements were refined using a global ultrasound elastography method, named global ultrasound elastography network (GLUENet) [85]. Recently, the same group proposed a modified PWC-Net (MPWC-Net) to obtain similar performance as state-of-the-art conventional elastography methods for liver strain imaging [86]. A thorough evaluation of CNN-based OF networks for breast ultrasound strain elastography can be found in [87].

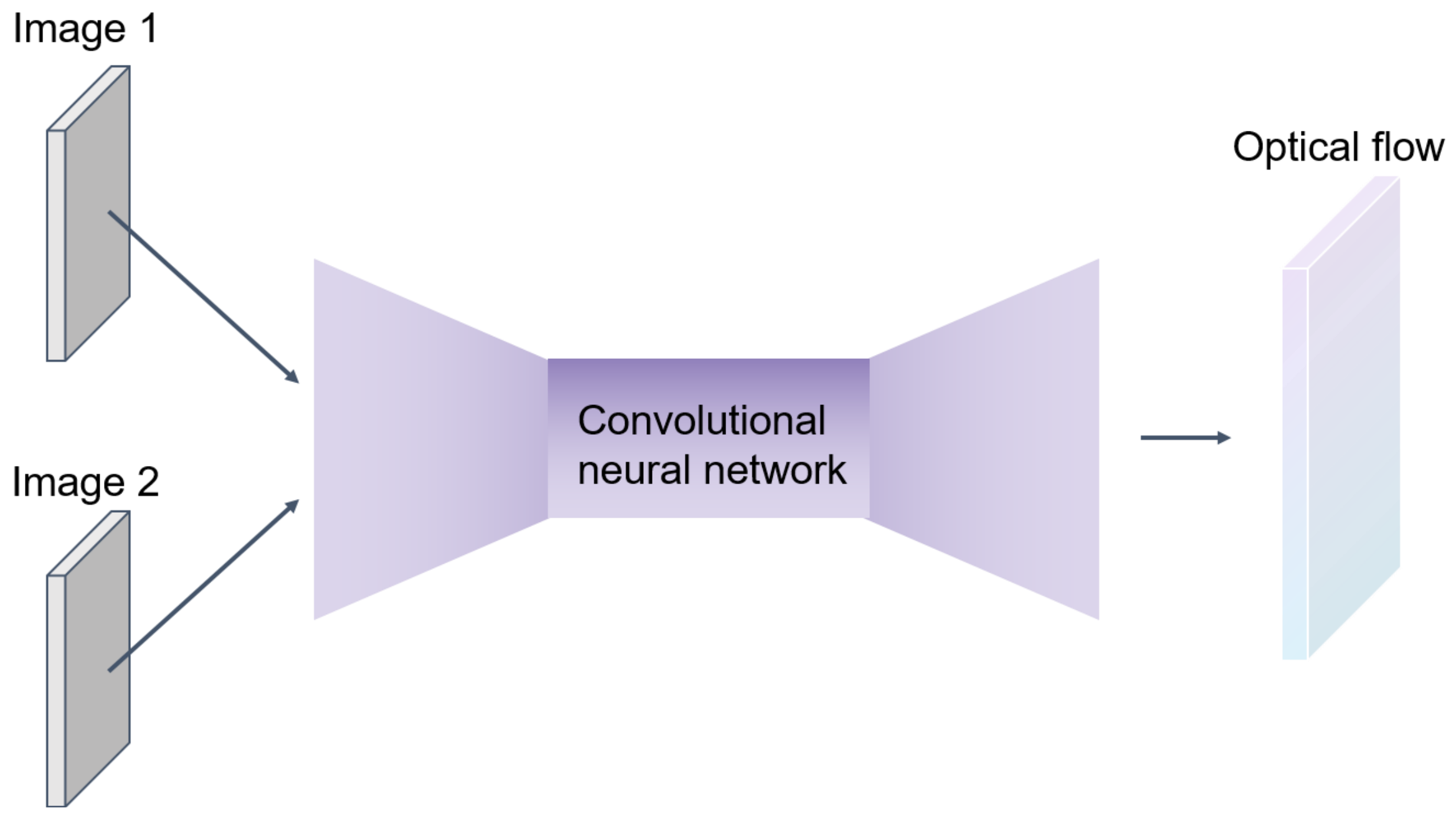

**Figure 6.** Architecture based on a neural network that learns to estimate end-to-end optical flow tracking of motions. The input is a pair of images and the output are optical flow estimations.

The straightforward translation from CNN-based OF architectures in computer vision to ultrasound quasi-static elastography is not adequate for accurate strain estimations. The main reason is that previous CNN-based networks were trained by datasets of photographs in which only rigid motions were represented [87]. However, tissue motions are more complex, including both rigid displacements and non-rigid deformations and rotations. Thus, other CNN-based networks, which are trained with ultrasound simulation datasets based on biomechanical models and ultrasound physics, have recently been proposed to provide end-to-end strain estimations. Wu et al. exploited two cascaded CNNs to reconstruct displacements and strains directly from RF signals [88]. A most recent network is shown in **Figure 7**, which incorporates a casual privileged information between the tissue displacement and strain. This learning-using-privileged-information (LUPI) paradigm corrects the intermediate state of the learning process [89], which allows the network to outperform state-of-the-art methods in terms of effectiveness and efficiency.

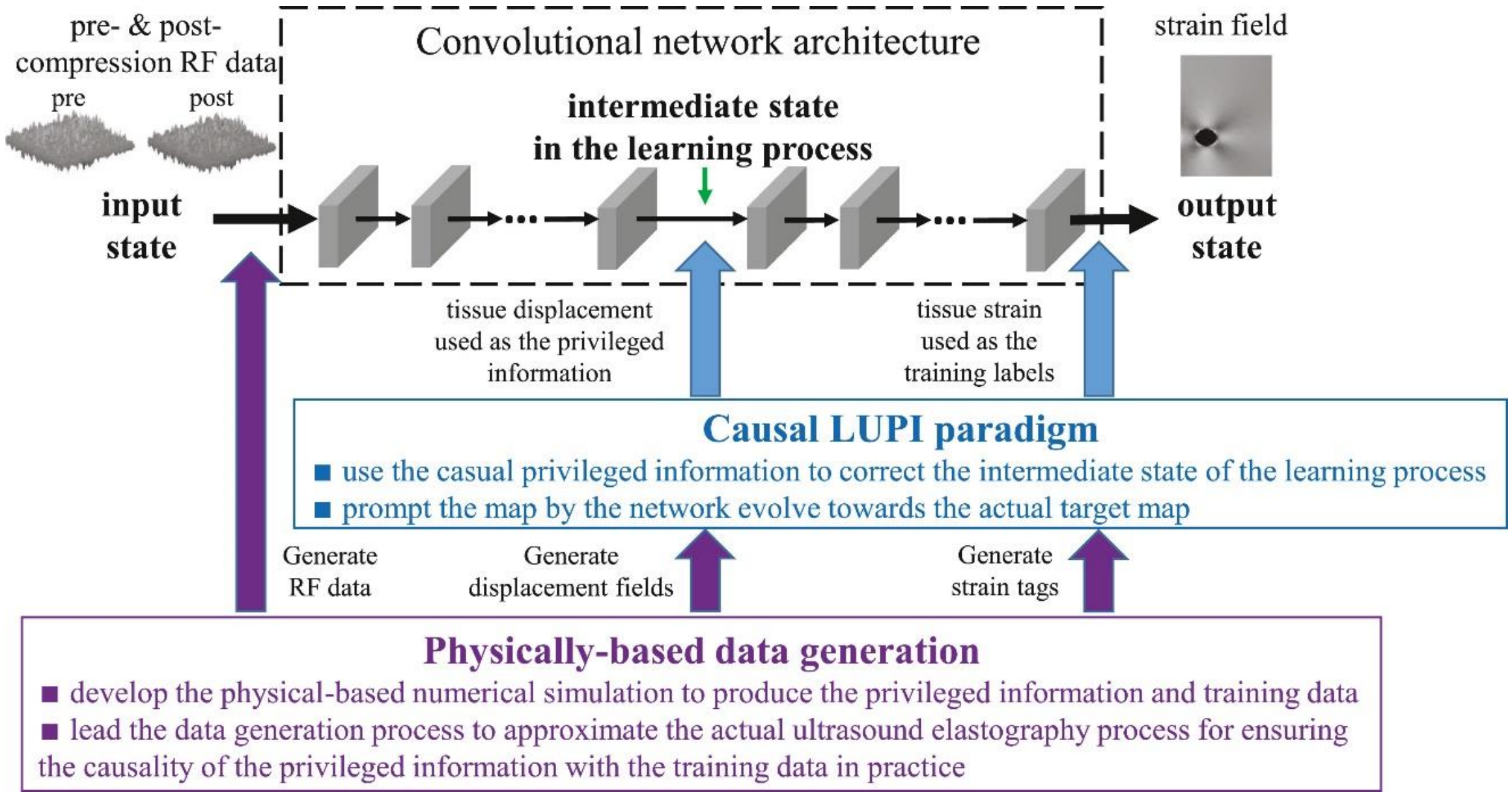

**Figure 7.** Overview of the LUPI network to reconstruct the strain field directly from pre- and post-compression RF signals. Reproduced with permission from [89]. Copyright 2019, Elsevier.

For myocardial elastography, FlowNet 2.0 was also incorporated to automatically assess myocardial function of the left ventricle in 2D echocardiography [90]. Since robust motion estimation is difficult for a left ventricle because of large translations, rotations and strains, and out-of-plane motions, some biomechanical prior assumptions such as spatiotemporal smoothness, are explicitly embedded in motion tracking methods to regularize motion fields and alleviate unreliable motion estimations. Lu et al. [91] proposed to use an MLP network that was trained with a synthetic cardiac dataset to learn a latent representation of spatiotemporal regularization, which showed an improvement in tracking performance.

### 4.2. Dynamic elastography applications

The development of deep neural networks in dynamic elastography has resulted in several clinical applications detailed next, which include classification of breast tumors [92, 93], liver imaging [94-96], thyroid nodules imaging [97], elastography of plantar fasciitis [98] and prostate cancer [99]. Such studies have been in the recent trends of the last decade. Architectures based on convolutional neural networks remain the most popular choices, while most studies have focused on either the classification of benign and malignant tumors, or on estimating the tissue stiffness. Let us review these studies while categorizing them according to their application.

#### 4.2.1 Applications in breast tumor classification

Shear wave elastography is being widely used for cancer screening, especially in breast cancer research. In general, malignant breast tissues are stiffer than healthy/benign ones, and morphological data can provide important clinical information. Ultrasound B-mode features have been widely used in conjunction with dynamic elastography data for breast tumor classification and the breast imaging reporting and data system (BI-RADS) has standardized the use of these approaches for diagnosis.

A shear wave elastography study based on a deep belief network of two-layer architecture comprising a point-wise Boltzmann machine and a restricted Boltzmann machine showed that deep learning features could achieve higher classification performance than conventional machine learning approaches with 93.4% accuracy, 88.6% sensitivity, and 97.1% specificity [92]. The continued development of deep learning showed great potential for classification, especially by utilizing a convolutional neural network architecture. A CNN based radiomics classification model for shear wave elastography data was first implemented by Zhou et al. [100]. The developed network could automatically extract a large number of features from the recorded dataset, including small structures/edges, as well as high level high-abstraction information. Their model reported as much as 4224 features extracted by the whole CNN architecture [100]. The first several layers of the CNN learned low level features that have small receptive fields such as edges, corners, etc… The middle layers learned to detect a part of the tumors. The top layers had larger receptive fields and could learn high-level, high-abstraction, and semantic features, such as parts, objects, etc… The only manual intervention required when using such architecture was selecting the ROI, which was done by experienced radiologists. There was no requirement for segmentation algorithms to extract crafted features such as contours, sizes, or shapes. **Figure 8** shows a schematic diagram of the CNN architecture that was used in the latter study for tumor classification.

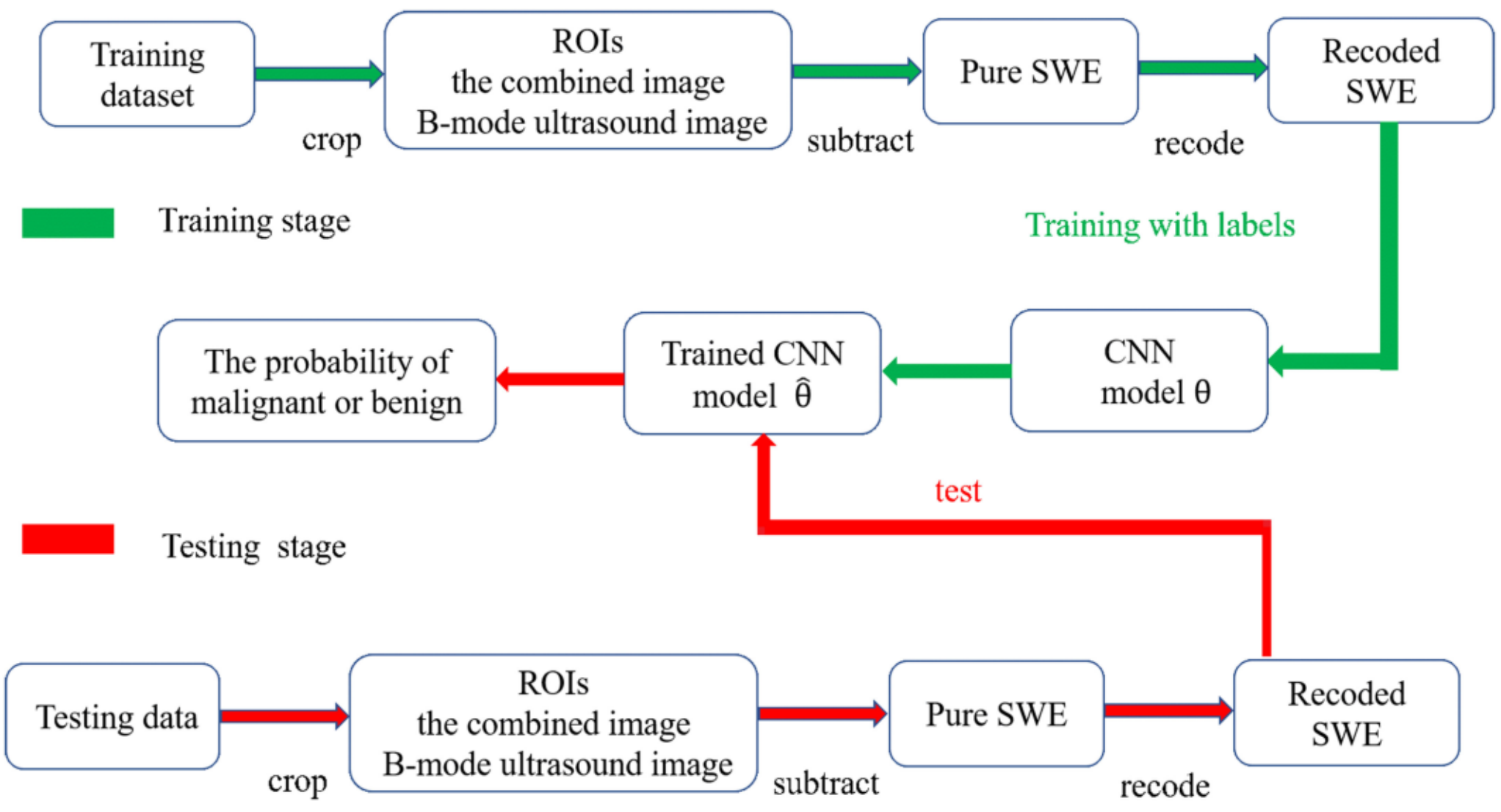

**Figure 8**. Schematic diagram of CNN for the classification of breast tumors on recoded SWE images. Reproduced with permission from [100]. Copyright 2018, IEEE.

Often in deep learning based studies, ultrasound elastography is utilized as a supporting parameter to other clinical ultrasound images to quantify a tumor grade by using a standardized color scheme. Elastography is probably more useful than color Doppler imaging for differentiating small, oval, or round triple-negative breast cancer from fibroadenoma [101]. A deep polynomial network artificial intelligence architecture for the automated extraction of dual-modal image features from both shear-wave elastography and B-mode ultrasound, with the leave-one-out cross validation, resulted in an impressive sensitivity of 97.8%, a specificity of 94.1%, and an accuracy of 95.6% [93]. Machine learning architectures based on support vector machines have also utilized tissue elasticity data for breast tumor segmentation [102].

Compared to other imaging modalities such as X-ray mammography or magnetic resonance imaging, AI architecture has been utilized less often in the ultrasound field in general and ultrasound elastography in particular. One of the main reasons is that poor image quality is observed when a large amount of tissue is examined by the ultrasound field, and shadowing of tumors makes their contours unclear. Also, in shear wave elastography, images can be misinterpreted if the probe is pressed harder [103]. Difficulty in data collection and recruiting a large number of patients, as mentioned earlier, are other reasons for fewer publications. The role of data augmentation, shown to be useful in avoiding overfitting and improving classification performance in various fields, has not been explored much in these studies on elastography and remains a low hanging fruit.

**4.2.2 Applications in liver imaging**

Chronic liver disease is a major health problem with most prevalently including non-alcoholic fatty liver disease (NAFLD) and viral hepatitis. Liver biopsy remains the common method for diagnosis, which is an invasive histologic reference standard with risk of complications. The liver imaging reporting and data system (LI-RADS) is the standardized classification system for imaging liver lesions and related diagnosis. Elastography for the assessment of liver stiffness by utilizing viscoelasticity biomarkers has replaced the need for a great majority of liver biopsies. However, an unequivocal advocacy of elastography as a complete replacement for a liver biopsy is still not established, and challenges in obtaining a perfect assessment of full-bodied patients with large body mass indexes, steatosis, or very stiff livers still need to be overcome [104-106]. Recent developments include deep networks that estimate liver fibrosis severity from ultrasound texture [107], deep learning radiomics of shear wave elastography [95], and the continued study of artificial intelligence (AI) based methods for medical imaging of the liver [94, 96]. Future objectives for the development of elastography systems include the diagnosis of inflammation, an important factor in the pathogenesis of non-alcoholic steatohepatitis (NASH) or viral hepatitis. A deeper understanding of the significance of the tissue mechanical properties as a standalone predictor of clinical outcomes is also desired. Elastography will continue to have rapid developments in the near future and artificial intelligence based diagnostic methods [94, 95, 107, 108] may have the potential to completely overturn the challenges faced in current state-of-the-art ultrasound B-mode and elastography imaging standards. Thus, AI studies in ultrasound B-mode liver imaging [109-112], as well as in elastography [95, 96], have accelerated in recent years.

One of the first AI models was a machine learning study based on a support vector machine model and a computer-aided diagnosis system for the classification of chronic liver disease using 2D-SWE imaging [113]. This study included a dataset of 126 subjects and a stiffness value clustering algorithm was employed to extract 35 features indicative of physical characteristics within SWE images. The study reported an accuracy of 87.3% with sensitivity and specificity values of 93.5% and 81.2%, respectively [113]. Wei et al. [114] proposed to reconstruct a clinical scoring system for improved detection of advanced hepatic fibrosis and cirrhosis in hepatitis B and C virus patients. Recently, deep models have become a more popular choice due to their better performance, as summarized next.

A multicenter study named deep learning radiomics of elastography (DLRE) was conducted by Wang et al. [95] to assess liver fibrosis stages on 398 patients from 12 hospitals with 1990 2D-SWE images. The patients had chronic hepatitis B and histology obtained from liver biopsy was used as a reference. They reported an area under the ROC curve (AUC) for cirrhosis of 0.97, for advanced fibrosis of 0.98, and for significant fibrosis of 0.85 with the use of the DLRE architecture. A schematic of the DLRE architecture is shown in **Figure 9**. Another study aimed at assessing chronic liver disease with multiple deep learning networks, including GoogLeNet, AlexNet, VGG16, testNet50, and DenseNet201, utilized shear wave elastography images acquired from 200 patients and registered an accuracy ranging from 87.2% - 97.4%, which outperformed the radiologists' performance at least by 10% [115]. Treacher et al. [107] focused on studying the texture pattern in grayscale elastography images to estimate shear wave velocity in liver fibrosis patients and investigated a total of 100 CNN architectures.. Gatos et al. [116] performed a study to automatically detect and isolate areas of low and high stiffness temporal stability in shear wave elastography images to define their impact in chronic liver disease diagnosis.

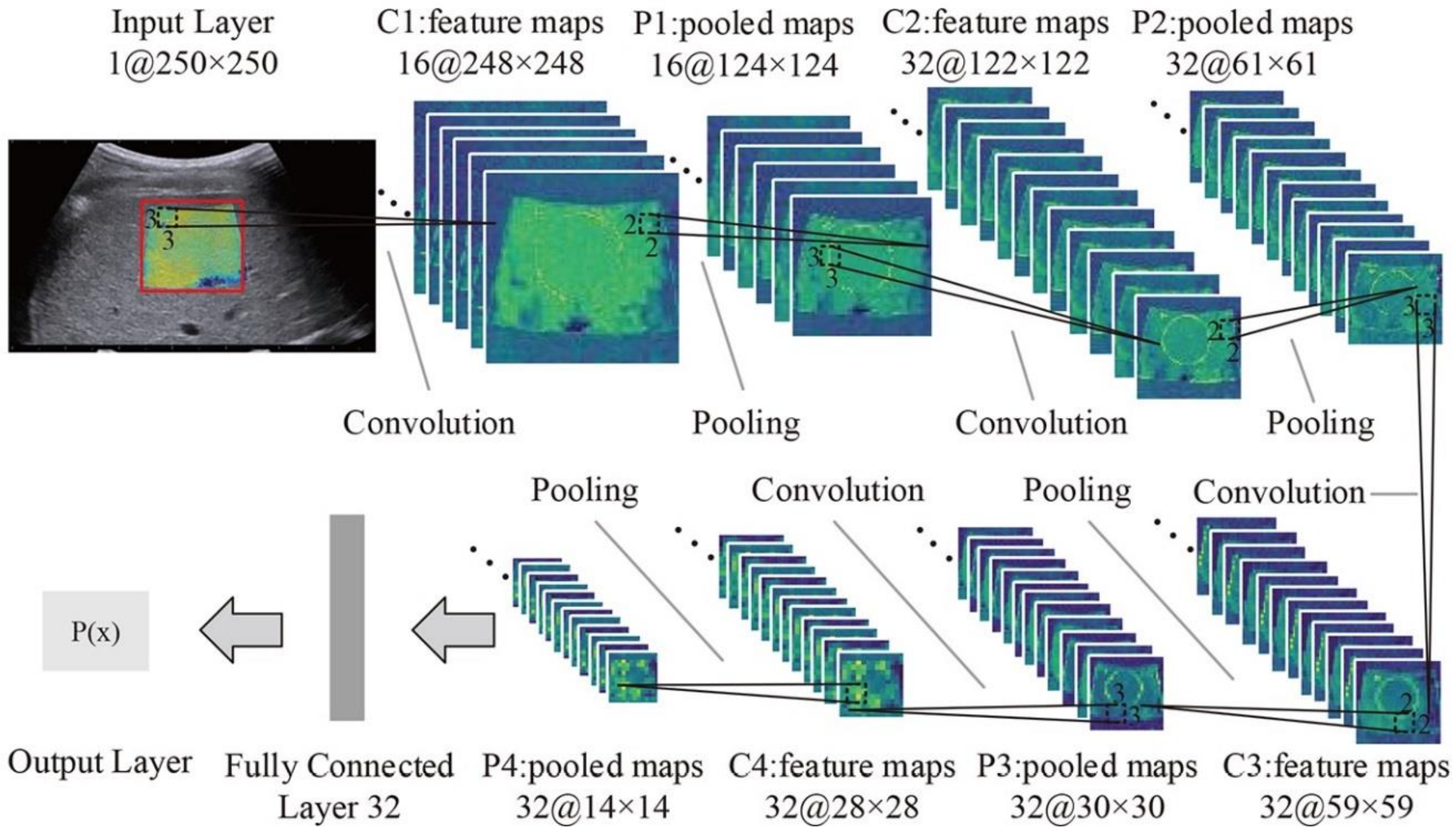

**Figure 9.** Illustration of the DLRE architecture that takes a 2D-SWE image as input to predict liver fibrosis stages. The input layer follows a four convolution-pooling procedure (C1-P1 to C4-P4), and the last pooled maps are fully connected with 32 neural nodes that perform classification. Reproduced with permission from [95]. Copyright 2019, British Medical Association.

Deep learning studies for liver imaging face the challenge that the acquired dataset is not of superior quality, often due to ultrasound attenuation in large body size. In addition, differences in imaging systems could affect the image display. Moreover, the imaging acquisition and examination need to be unified among operators to construct a well-qualified imaging database. The implementation of transfer learning strategies and further deep learning advancements could improve classification accuracy.

### 4.2.3 Applications in thyroid nodule classification

Thyroid nodules are abnormal lumps in the organ, which may be indicators of cancer. As for other organs, the standardized diagnostic algorithm for thyroid imaging is the reporting and data system (TI-RADS), which provides a score to indicate the risk of malignancy of the thyroid gland. Confirmation of diagnosis still requires fine needle inspiration biopsy; however due to its complexity, about 10%–20% of thyroid nodule biopsies tend to be non-diagnostic [117]. Qin et al. [97] utilized a combined B-mode and elasticity image dataset as an input to a CNN architecture for combining depth features of these images to form a hybrid space for the classification of benign and malignant thyroid nodules. As far as we know, no other AI studies have been reported.

### 4.2.4 Other clinical applications

In addition to the three main applications discussed here, AI based diagnostic research is being carried out for several other clinical applications. One of the first such applications was in endoscopic ultrasound elastography for the diagnosis of focal pancreatic masses where a multilayer perceptron architecture was used [118]. In 2D shear wave elastography, Gao et al. [98] developed a deep Siamese framework with multitask learning (DS-MLTL) and transfer learning, which could learn discriminative visual features and effective recognition functions for the classification of plantar fasciitis, a common cause of heel pain. The study also reported that their network resulted in higher sensitivity and specificity compared to other popular deep learning methods, namely the CNN model, transfer learning model, and deep Siamese multitask learning model. The table below provides a summary of technical informations on different clinical applications. As seen, various AI models were used and training and test sets were limited with less than 1500 images from below 270 patients.

**Table 1. A summary of clinical diagnostic applications using dynamic elastography and deep learning methods**

| Disease/ application | AI classifier | Training dataset (images) | Testing dataset (images) | Accuracy (%) | Sensitivity/ Specificity (%) |
|---|---|---|---|---|---|
| Breast tumors [93] | Deep polynomial network | 227 dual mode (SWE and B-mode ultrasound) from 121 patients | | 95.6 | 97.8/94.1 |
| Breast tumors [100] | CNN | 400 (training) and 140 (testing) from 205 patients | | 95.8 | 96.2/95.7 |
| Breast tumors [92] | Unified point-wise gated Boltzmann machine based two-layer deep learning architecture | 227 | 5 | 93.4 | 88.6/97.1 |
| Chronic liver disease [115] | CNN | 140 | 60 | 87.2-97.4 | - |
| Hepatitis B virus-associated fibrosis [95] | CNN | 1330 from 266 patients | 660 from 132 patients | 0.98 (AUC)[(a)] | 90.4/98.3 |
| Liver fibrosis [116] | CNN | 200 from 200 patients | | 82.5-95.5 | - |
| Thyroid nodule [97] | CNN | 908 from 183 patients | 248 from 40 patients | 94.7 | 92.7/+97.9 |
| Plantar fasciitis [98] | Deep siamese framework | 720 | 360 | 85.0 | 76.8/93.3 |

[(a)] Area under the receiver-operating curve

It is worth mentioning that artificial intelligence architectures have also been utilized in several system development related applications that could be used in elastography, such as beamforming [119] and inverse problem solution in elasticity imaging [120]. The scope for both deep learning based advanced architectures seems to be endless in medical diagnostics. It is expected that elastography data in combination with conventional ultrasound data could greatly improve the diagnostic accuracy and advantages are multifold. Rightly so, deep learning has been dubbed as a paradigm shift in medical imaging [121] that could achieve optimal results cost-effectively, especially from large and compromised datasets, as well as for non-linear, non-convex, and overly complex problems.

## 5. Conclusions and perspectives

Deep neural networks provide automated solutions for several clinical problems that may require expert skills for humans. There are two major fields of clinical AI research: (a) image processing/analysis which involves denoising, feature extraction from images, segmentation etc…, and (b) image formation/reconstruction i.e., from data to images. Even though a deep network is fundamentally different from multi-resolution analysis or optimization models, the major virtue of deep networks is the non-linear learning and non-convex optimization ability. However, the lack of curated training data is a major limitation in advancing any AI model. The time needed to translate fully automated deep models for diagnosis tasks with acceptance by the clinical community may range from a few years to decades. There is a need to develop more advanced algorithms to solve more complex medical problems, especially in ultrasound imaging. There is also a need to validate diagnostic assistance methods (e.g., heatmaps [122]) providing feedback to clinicians instead of final diagnosis.

This article summarized several studies that implemented deep learning methodologies for ultrasound elastography applications, particularly for image analysis and feature extraction. In the next paragraphs, a discussion is given on some prospective topics of research that have not yet been explored much in ultrasound elastography AI research.

There has been relatively limited research conducted on image reconstruction in shear wave elastography applications, which remains a low hanging fruit [123]. Deep learning based reconstruction algorithms have data driven knowledge-enhancing abilities that provide smarter initial guess, more relevant features extraction, and application-specific regularized final images. If a dataset is severely compromised due to truncation, distortion, noise, or artifacts, a model-based iterative or analytical algorithm can be used for initial reconstruction and subsequently used as an input to deep network [124]. This two-step process for image reconstruction can be extremely useful in elastography applications. Such networks can become favorable since deep networks with images as inputs can easily be adapted and domain-specific data can be incorporated as unprecedented prior knowledge. Bandwidth enhancement of recorded data is another open field [125] that can improve reconstruction algorithms in ultrasound elastography.

The network design remains a popular topic of research in terms of both overall architecture and component characteristics [92, 93, 126, 127]. This is equivalent to algorithmic design or computer architecture design, which still is a prominent topic for specific applications such as elastography. Another important field is unsupervised learning that includes generative adversarial networks, which can realize automated data curation by learning discriminatory features from an unlabeled dataset. Several studies have explored unsupervised learning in magnetic resonance and computed tomography imaging modalities and similar exploration is needed in ultrasound elastography applications. Artificial intelligence research can be further improved by mimicking neuroplasticity – the ability of the brain to reorganize itself by following new neural connections for better learning, adaption, and compensation.

Finally, deep models also have some limitations. For example, deep learning methods do not seem to be applicable widely in clinical practice, especially in ultrasound elastography, due to variability caused by the data itself. Such models may suffer from poor generalization that comes from different machines with different acquisition parameters. Another challenge is that these models are usually trained for only one task. Thus, it is almost impossible for AI to replace radiologists in the coming decades, however, radiologists who can use AI will ultimately replace conventional radiologists [94, 128].

Rigorous data collection criteria and uniform global ethical guidelines for AI need to be developed in order to establish its role in clinical practice and for advancing applications in elastography. Ethical and legal responsibilities for decision making will remain under the responsibility of physicians and AI should adapt to this context. Although it has been well demonstrated that humans and AI models working together can achieve a higher level of accuracy in ultrasound elastography based diagnosis and prognosis, efforts, legislation strategies and legal issues will have to be addressed to include AI into clinical guideline practices.

One common obstacle across research groups is that it is not easy for single sites to generate a large amount of data and to label it. Multicenter initiatives, collaborative research, data sharing, and crowd sourcing are some possible approaches to unite useful resources for further deep learning based advancements in elastography. Scientists must consider sharing clinical data for further improvement in this field. Zhou et al [94] also called for creating a globally interconnected network of an open patient dataset to include diverse demographics and geographical locations. This would also promote AI research in countries where acquiring patient datasets is a challenge due to ethical, financial, or bureaucratic limitations. Artificial intelligence in medical research will only move forward in a socially responsible and globally beneficial way by promoting open access research and shared clinical datasets.

**Key Points**

- Artificial intelligence models trained on multiple modalities (B-mode, shear wave elastography, Doppler ultrasound, etc…) could improve diagnostic accuracy multifold.
- Convolutional neural network architectures are the most popular for classification applications, especially in breast cancer screening and liver fibrosis grading.
- There is a greater need for datasets that are less operator and machine dependent. Providing unified guidance to operators would make ultrasound and shear wave elastography more intelligent.
- Deep learning based diagnostic approaches are more efficient than traditional machine learning approaches.

**Acknowledgements**
H. L. acknowledges the fellowship support of the TransMedTech Institute. This work was funded by the Canadian Institutes of Health Research (no. 399544, 273738, 301520 and 134748), Fonds de Recherche Québec (no. 2019-AUDC-263591) and Collaborative Health Research Program of the Natural Sciences and Engineering Research Council of Canada (no. 462240-14).

**Conflict of Interest**
The authors declare no conflict of interest.

References

[1] A. P. Sarvazyan, A. R. Skovoroda, S. Y. Emelianov, J. B. Fowlkes, J. G. Pipe, R. S. Adler, R. B. Buxton, and P. L. Carson, "Biophysical bases of elasticity imaging," in *Acoustical Imaging*: Springer, 1995, pp. 223-240.

[2] M. A. Meyers and K. K. Chawla, *Mechanical behavior of materials*. New York: The Cambridge University Press, 2008.

[3] A. S. Saada, *Elasticity: theory and applications*. Elsevier, 2013.

[4] J. F. Greenleaf, M. Fatemi, and M. Insana, "Selected methods for imaging elastic properties of biological tissues," *Annu Rev Biomed Eng,* vol. 5, pp. 57-78, 2003.

[5] K. J. Parker, M. M. Doyley, and D. J. Rubens, "Imaging the elastic properties of tissue: the 20 year perspective," *Physics in Medicine and Biology,* vol. 57, no. 16, pp. 5359-5360, 2012.

[6] S. Chen, M. Fatemi, and J. F. Greenleaf, "Remote measurement of material properties from radiation force induced vibration of an embedded sphere," *J Acoust Soc Am,* vol. 112, no. 3, pp. 884-889, Sep 2002.

[7] J. Ophir, I. Cespedes, H. Ponnekanti, Y. Yazdi, and X. Li, "Elastography: a quantitative method for imaging the elasticity of biological tissues," *Ultrasonic imaging,* vol. 13, no. 2, pp. 111-134, 1991.

[8] T. Shiina, K. R. Nightingale, M. L. Palmeri, T. J. Hall, J. C. Bamber, R. G. Barr, L. Castera, B. I. Choi, Y. H. Chou, D. Cosgrove, C. F. Dietrich, H. Ding, D. Amy, A. Farrokh, G. Ferraioli, C. Filice, M. Friedrich-Rust, K. Nakashima, F. Schafer, I. Sporea, S. Suzuki, S. Wilson, and M. Kudo, "WFUMB guidelines and recommendations for clinical use of ultrasound elastography: Part 1: basic principles and terminology," *Ultrasound Med Biol,* vol. 41, no. 5, pp. 1126-1147, May 2015.

[9] H. Hansen, R. Lopata, and C. L. de Korte, "Noninvasive carotid strain imaging using angular compounding at large beam steered angles: Validation in vessel phantoms," *IEEE Transactions on Medical Imaging,* vol. 28, no. 6, pp. 872-880, 2009.

[10] R. G. Lopata, M. M. Nillesen, J. M. Thijssen, L. Kapusta, and C. L. de Korte, "Three-dimensional cardiac strain imaging in healthy children using RF-data," *Ultrasound Med Biol,* vol. 37, no. 9, pp. 1399-1408, Sep 2011.

[11] E. A. Bunting, J. Provost, and E. E. Konofagou, "Stochastic precision analysis of 2D cardiac strain estimation in vivo," *Phys Med Biol,* vol. 59, no. 22, pp. 6841-6858, Nov 2014.

[12] R. L. Maurice, J. Ohayon, Y. Fretigny, M. Bertrand, G. Soulez, and G. Cloutier, "Noninvasive vascular elastography: theoretical framework," *IEEE Transactions on Medical Imaging,* vol. 23, no. 2, pp. 164-180, 2004.

[13] J. Poree, D. Garcia, B. Chayer, J. Ohayon, and G. Cloutier, "Noninvasive vascular elastography with plane strain incompressibility assumption using ultrafast coherent compound plane wave imaging," *IEEE Trans Med Imaging,* vol. 34, no. 12, pp. 2618-2631, Dec 2015.

[14] H. Li, J. Porée, B. Chayer, M.-H. R. Cardinal, and G. Cloutier, "Parameterized strain estimation for vascular ultrasound elastography with sparse representation," *IEEE Transactions on Medical Imaging,* in press June 2020.

[15] S. A. McAleavey, M. Menon, and J. Orszulak, "Shear-modulus estimation by application of spatially-modulated impulsive acoustic radiation force," *Ultrasonic imaging,* vol. 29, no. 2, pp. 87-104, 2007.

[16] H. L. Oestreicher, "Field and impedance of an oscillating sphere in a viscoelastic medium with an application to biophysics," *J Acoust Soc Am,* vol. 23, no. 6, pp. 707-714, Jun 1951.

[17] J. Ophir, S. K. Alam, B. Garra, F. Kallel, E. Konofagou, T. Krouskop, and T. Varghese, "Elastography: ultrasonic estimation and imaging of the elastic properties of tissues," *Proceedings of the Institution of Mechanical Engineers, Part H: Journal of Engineering in Medicine,* vol. 213, no. 3, pp. 203-233, 1999.
[18] S. Kazemirad, S. Bernard, S. Hybois, A. Tang, and G. Cloutier, "Ultrasound shear wave viscoelastography: model-independent quantification of the complex shear modulus," *IEEE Trans Ultrason Ferroelectr Freq Control,* vol. 63, no. 9, pp. 1399-1408, Sep 2016.
[19] A. P. Sarvazyan, O. V. Rudenko, S. D. Swanson, J. B. Fowlkes, and S. Y. Emelianov, "Shear wave elasticity imaging: a new ultrasonic technology of medical diagnostics," *Ultrasound in medicine & biology,* vol. 24, no. 9, pp. 1419-1435, 1998.
[20] Z. Qu and Y. Ono, "A method to reduce the influence of reflected waves on shear velocity measurements using B-mode scanning time delay," *Japanese Journal of Applied Physics,* vol. 54, no. 7S1, 2015.
[21] M. Fatemi and J. F. Greenlea, "Ultrasound-stimulated vibro-acoustic spectrography," *Science,* vol. 280, no. 5360, pp. 82-85, 1998.
[22] S. Chen, M. W. Urban, C. Pislaru, R. Kinnick, Y. Zheng, A. Yao, and J. F. Greenleaf, "Shearwave dispersion ultrasound vibrometry (SDUV) for measuring tissue elasticity and viscosity," *IEEE transactions on ultrasonics, ferroelectrics, frequency control,* vol. 56, no. 1, pp. 55-62, 2009.
[23] L. Gao, K. J. Parker, R. M. Lerner, and S. F. Levinson, "Imaging of the elastic properties of tissue—A review," *Ultrasound in medicine & biology,* vol. 22, no. 8, pp. 959-977, 1996.
[24] H. Eskandari, S. E. Salcudean, and R. Rohling, "Viscoelastic parameter estimation based on spectral analysis," *IEEE Trans Ultrason Ferroelectr Freq Control,* vol. 55, no. 7, pp. 1611-1125, Jul 2008.
[25] E. Turgay, S. Salcudean, and R. Rohling, "Identifying the mechanical properties of tissue by ultrasound strain imaging," *Ultrasound Med Biol,* vol. 32, no. 2, pp. 221-235, Feb 2006.
[26] M. Sridhar, J. Liu, and M. F. Insana, "Viscoelasticity imaging using ultrasound: parameters and error analysis," *Phys Med Biol,* vol. 52, no. 9, pp. 2425-2443, May 7 2007.
[27] E. E. Konofagou, J. D'hooge, and J. Ophir, "Myocardial elastography—A feasibility study in vivo," *Ultrasound in medicine & biology,* vol. 28, no. 4, pp. 475-482, 2002.
[28] C. L. de Korte, A. F. van der Steen, E. I. Céspedes, G. Pasterkamp, S. G. Carlier, F. Mastik, A. H. Schoneveld, P. W. Serruys, and N. Bom, "Characterization of plaque components andvulnerability with intravascular ultrasound elastography," *Physics in Medicine & Biology,* vol. 45, no. 6, pp. 1465-1475, 2000.
[29] C. L. de Korte, S. G. Carlier, F. Mastik, M. M. Doyley, A. F. van der Steen, P. W. Serruys, and N. Bom, "Morphological and mechanical information of coronary arteries obtained with intravascular elastography: feasibility study in vivo," *Eur Heart J,* vol. 23, no. 5, pp. 405-413, Mar 2002.
[30] H. Kanai, H. Hasegawa, M. Ichiki, F. Tezuka, and Y. Koiwa, "Elasticity imaging of atheroma with transcutaneous ultrasound: preliminary study," *Circulation,* vol. 107, no. 24, pp. 3018-3021, Jun 24 2003.
[31] S. Korukonda and M. M. Doyley, "Visualizing the radial and circumferential strain distribution within vessel phantoms using synthetic-aperture ultrasound elastography," *IEEE Trans Ultrason Ferroelectr Freq Control,* vol. 59, no. 8, pp. 1639-1653, Aug 2012.
[32] H. Li, B. Chayer, M.-H. Roy Cardinal, J. Muijsers, M. van den Hoven, Z. Qin, M. Gesnik, G. Soulez, R. G. P. Lopata, and G. Cloutier, "Investigation of out-of-plane

motion artifacts in 2D noninvasive vascular ultrasound elastography," *Physics in Medicine & Biology*, vol. 63, no. 24, p. 245003, 2018.
[33] M. E. Anderson, "Multi-dimensional velocity estimation with ultrasound using spatial quadrature," *IEEE Transactions on Ultrasonics, Ferroelectrics and Frequency Control*, vol. 45, pp. 852-861, 1998.
[34] J. A. Jensen and P. Munk, "A new method for estimation of velocity vectors," *IEEE Transactions on Ultrasonics, Ferroelectrics, and Frequency Control*, vol. 45, pp. 886-894, 1998.
[35] H. Li, J. Poree, M. H. Roy Cardinal, and G. Cloutier, "Two-dimensional affine model-based estimators for principal strain vascular ultrasound elastography with compound plane wave and transverse oscillation beamforming," *Ultrasonics*, vol. 91, pp. 77-91, Jan 2019.
[36] R. M. Lerner, K. J. Parker, J. Holen, R. Gramiak, and R. C. Waag, "Sono-elasticity: Medical elasticity images derived from ultrasound signals in mechanically vibrated targets," in *Acoustical imaging*: Springer, 1988, pp. 317-327.
[37] Y. Yamakoshi, J. Sato, and T. Sato, "Ultrasonic imaging of internal vibration of soft tissue under forced vibration," *IEEE transactions on ultrasonics, ferroelectrics, and frequency control*, vol. 37, no. 2, pp. 45-53, 1990.
[38] S. Catheline, F. Wu, and M. Fink, "A solution to diffraction biases in sonoelasticity: the acoustic impulse technique," *J Acoust Soc Am*, vol. 105, no. 5, pp. 2941-2950, May 1999.
[39] J. Bercoff, M. Tanter, and M. Fink, "Supersonic shear imaging: a new technique for soft tissue elasticity mapping," *IEEE transactions on ultrasonics, ferroelectrics, and frequency control*, vol. 51, no. 4, pp. 396-409, 2004.
[40] K. R. Nightingale, M. L. Palmeri, R. W. Nightingale, and G. E. Trahey, "On the feasibility of remote palpation using acoustic radiation force," *J Acoust Soc Am*, vol. 110, no. 1, pp. 625-634, Jul 2001.
[41] P. Song, M. C. Macdonald, R. H. Behler, J. D. Lanning, M. H. Wang, M. W. Urban, A. Manduca, H. Zhao, M. R. Callstrom, A. Alizad, J. F. Greenleaf, and S. Chen, "Two-dimensional shear-wave elastography on conventional ultrasound scanners with time-aligned sequential tracking (TAST) and comb-push ultrasound shear elastography (CUSE)," *IEEE transactions on ultrasonics, ferroelectrics, frequency control*, vol. 62, no. 2, pp. 290-302, 2015.
[42] T. Deffieux, J.-L. Gennisson, L. Bousquet, M. Corouge, S. Cosconea, D. Amroun, S. Tripon, B. Terris, V. Mallet, P. Sogni, M. Tanter, and S. Pol, "Investigating liver stiffness and viscosity for fibrosis, steatosis and activity staging using shear wave elastography," *Journal of hepatology*, vol. 62, no. 2, pp. 317-324, 2015.
[43] A. E. Samir, M. Dhyani, A. Vij, A. K. Bhan, E. F. Halpern, J. Méndez-Navarro, K. E. Corey, and R. T. Chung, "Shear-wave elastography for the estimation of liver fibrosis in chronic liver disease: determining accuracy and ideal site for measurement," *Radiology*, vol. 274, no. 3, pp. 888-896, 2015.
[44] A. Goddi, M. Bonardi, and S. Alessi, "Breast elastography: A literature review," *J Ultrasound*, vol. 15, no. 3, pp. 192-198, Sep 2012.
[45] A. Evans, P. Whelehan, K. Thomson, D. McLean, K. Brauer, C. Purdie, L. Baker, L. Jordan, P. Rauchhaus, and A. Thompson, "Invasive breast cancer: relationship between shear-wave elastographic findings and histologic prognostic factors," *Radiology*, vol. 263, no. 3, pp. 673-677, 2012.
[46] R. G. Barr, R. Memo, and C. R. Schaub, "Shear wave ultrasound elastography of the prostate: initial results," *Ultrasound quarterly*, vol. 28, no. 1, pp. 13-20, 2012.
[47] F. Sebag, J. Vaillant-Lombard, J. Berbis, V. Griset, J. F. Henry, P. Petit, and C. Oliver, "Shear wave elastography: a new ultrasound imaging mode for the differential

diagnosis of benign and malignant thyroid nodules," *J Clin Endocrinol Metab,* vol. 95, no. 12, pp. 5281-5288, Dec 2010.
[48] L. C. Davis, T. G. Baumer, M. J. Bey, and M. V. Holsbeeck, "Clinical utilization of shear wave elastography in the musculoskeletal system," *Ultrasonography,* vol. 38, no. 1, pp. 2-12, Jan 2019.
[49] J. Brum, M. Bernal, J. L. Gennisson, and M. Tanter, "In vivo evaluation of the elastic anisotropy of the human Achilles tendon using shear wave dispersion analysis," *Phys Med Biol,* vol. 59, no. 3, pp. 505-523, Feb 7 2014.
[50] K. S. Bhatia, C. C. Cho, C. S. Tong, E. H. Yuen, and A. T. Ahuja, "Shear wave elasticity imaging of cervical lymph nodes," *Ultrasound Med Biol,* vol. 38, no. 2, pp. 195-201, Feb 2012.
[51] C. Jenssen and C. F. Dietrich, "Endoscopic ultrasound-guided fine-needle aspiration biopsy and trucut biopsy in gastroenterology - An overview," *Best Pract Res Clin Gastroenterol,* vol. 23, no. 5, pp. 743-759, 2009.
[52] S. Paterson, F. Duthie, and A. J. Stanley, "Endoscopic ultrasound-guided elastography in the nodal staging of oesophageal cancer," *World J Gastroenterol,* vol. 18, no. 9, pp. 889-895, Mar 2012.
[53] D. K. Teng, H. Wang, Y. Q. Lin, G. Q. Sui, F. Guo, and L. N. Sun, "Value of ultrasound elastography in assessment of enlarged cervical lymph nodes," *Asian Pac J Cancer Prev,* vol. 13, no. 5, pp. 2081-2085, 2012.
[54] A. E. Samir, A. S. Allegretti, Q. Zhu, M. Dhyani, A. Anvari, D. A. Sullivan, C. A. Trottier, S. Dougherty, W. W. Williams, J. L. Babitt, J. Wenger, R. I. Thadhani, and H. Y. Lin, "Shear wave elastography in chronic kidney disease: a pilot experience in native kidneys," *BMC Nephrol,* vol. 16, p. 119, Jul 2015.
[55] S. Aristizabal, C. Amador Carrascal, I. Z. Nenadic, J. F. Greenleaf, and M. W. Urban, "Application of acoustoelasticity to evaluate nonlinear modulus in ex vivo kidneys," *IEEE Trans Ultrason Ferroelectr Freq Control,* vol. 65, no. 2, pp. 188-200, Feb 2018.
[56] A. Caenen, M. Pernot, M. Peirlinck, L. Mertens, A. Swillens, and P. Segers, "An in silico framework to analyze the anisotropic shear wave mechanics in cardiac shear wave elastography," *Phys Med Biol,* vol. 63, no. 7, p. 075005, Mar 23 2018.
[57] M. Bhatt, M. A. C. Moussu, B. Chayer, F. Destrempes, M. Gesnik, L. Allard, A. Tang, and G. Cloutier, "Reconstruction of Viscosity Maps in Ultrasound Shear Wave Elastography," *IEEE Transactions on Ultrasonics, Ferroelectrics and Frequency Control,* vol. 66, no. 6, pp. 1065-1078, Apr 11 2019.
[58] G. Rus, I. H. Faris, J. Torres, A. Callejas, and J. Melchor, "Why are viscosity and nonlinearity bound to make an impact in clinical elastographic diagnosis?," *Sensors,* vol. 20, no. 8, p. 2379, Apr 22 2020.
[59] J. J. Fenton, S. H. Taplin, P. A. Carney, L. Abraham, E. A. Sickles, C. D’Orsi, E. A. Berns, G. Cutter, R. E. Hendrick, W. E. Barlow, and J. G. Elmore, "Influence of computer-aided detection on performance of screening mammography," *New England Journal of Medicine,* vol. 356, no. 14, pp. 1399-1409, 2007.
[60] G. Litjens, T. Kooi, B. E. Bejnordi, A. A. A. Setio, F. Ciompi, M. Ghafoorian, J. van der Laak, B. van Ginneken, and C. I. Sanchez, "A survey on deep learning in medical image analysis," *Med Image Anal,* vol. 42, pp. 60-88, Dec 2017.
[61] A. Krizhevsky, I. Sutskever, and G. E. Hinton, "Imagenet classification with deep convolutional neural networks," in *Advances in neural information processing systems*, 2012, pp. 1097-1105.
[62] S. Hochreiter and J. Schmidhuber, "Long short-term memory," *Neural computation,* vol. 9, no. 8, pp. 1735-1780, 1997.

[63] I. Sutskever, O. Vinyals, and Q. V. Le, "Sequence to sequence learning with neural network," in *Advances in neural information processing systems*, 2014, pp. 3104-3112.
[64] I. J. Goodfellow, J. Pouget-Abadie, M. Mirza, B. Xu, D. Warde-Farley, S. Ozair, A. Courville, and Y. Bengio, "Generative adversarial nets," in *Advances in neural information processing systems*, 2014, pp. 2672-2680.
[65] J.-Y. Zhu, T. Park, P. Isola, and A. A. Efros, "Unpaired image-to-image translation using cycle-consistent adversarial networks," in *IEEE international conference on computer vision*, 2017, pp. 2223-2232.
[66] J. Ker, L. Wang, J. Rao, and T. Lim, "Deep learning applications in medical image analysis," *IEEE Access*, vol. 6, pp. 9375-9389, 2017.
[67] J. Chen, L. Yang, Y. Zhang, M. Alber, and D. Z. Chen, "Combining fully convolutional and recurrent neural networks for 3d biomedical image segmentation," in *Advances in neural information processing systems*, 2016, pp. 3036-3044.
[68] S. Azizi, S. Bayat, P. Yan, A. Tahmasebi, J. T. Kwak, S. Xu, B. Turkbey, P. Choyke, P. Pinto, B. Wood, P. Mousavi, and P. Abolmaesumi, "Deep recurrent neural networks for prostate cancer detection: analysis of temporal enhanced ultrasound," *IEEE Trans Med Imaging*, vol. 37, no. 12, pp. 2695-2703, Dec 2018.
[69] S. Liu, Y. Wang, X. Yang, B. Lei, L. Liu, S. X. Li, D. Ni, and T. Wang, "Deep learning in medical ultrasound analysis: a review," *Engineering*, vol. 5, no. 2, pp. 261-275, 2019.
[70] I. Sutskever, J. Martens, G. Dahl, and G. Hinton, "On the importance of initialization and momentum in deep learning," in *International conference on machine learning*, 2013, pp. 1139-1147.
[71] D. P. Kingma and J. L. Ba, "Adam: A method for stochastic optimization," in *International Conference on Learning Representations*, 2015.
[72] V. Nair and G. E. Hinton, "Rectified linear units improve restricted boltzmann machines," in *International conference on machine learning* 2010, pp. 807-814.
[73] I. J. Goodfellow, D. Warde-Farley, M. Mirza, A. Courville, and Y. Bengio, "Maxout networks," in *International Conference on Machine Learning*, 2013, pp. 1319-1327.
[74] N. Srivastava, G. Hinton, A. Krizhevsky, I. Sutskever, and R. Salakhutdinov, "Dropout: a simple way to prevent neural networks from overfitting," *The journal of machine learning research*, vol. 15, no. 1, pp. 1929-1958, 2014.
[75] S. Ioffe and C. Szegedy, "Batch normalization: Accelerating deep network training by reducing internal covariate shift," in *International Conference on Machine Learning*, 2015, pp. 448–456.
[76] S. J. Pan and Q. Yang, "A survey on transfer learning," *IEEE Transactions on knowledge and data engineering*, vol. 22, no. 10, pp. 1345-1359, 2009.
[77] N. Tajbakhsh, J. Y. Shin, S. R. Gurudu, R. T. Hurst, C. B. Kendall, M. B. Gotway, and L. Jianming, "Convolutional neural networks for medical image analysis: full training or fine tuning?," *IEEE Trans Med Imaging*, vol. 35, no. 5, pp. 1299-1312, May 2016.
[78] H. Ravishankar, P. Sudhakar, R. Venkataramani, S. Thiruvenkadam, P. Annangi, N. Babu, and V. Vaidya, "Understanding the mechanisms of deep transfer learning for medical images," in *Deep learning and data labeling for medical applications*: Springer, 2016, pp. 188-196.
[79] P. Fischer, A. Dosovitskiy, E. Ilg, P. Häusser, C. Hazırbas, V. Golkov, P. v. d. Smagt, D. Cremers, and T. Brox, "FlowNet: learning optical flow with convolutional networks," *Proceedings of the IEEE international conference on computer vision*, pp. 2758-2766, 2015.

[80] E. Ilg, N. Mayer, T. Saikia, M. Keuper, A. Dosovitskiy, and T. Brox, "FlowNet 2.0: Evolution of optical flow estimation with deep networks," *Proceedings of the IEEE conference on computer vision and pattern recognition,* pp. 2462-2470, 2017.
[81] A. Ranjan and M. J. Black, "Optical flow estimation using a spatial pyramid network," *Proceedings of the IEEE Conference on Computer Vision and Pattern Recognition,* pp. 4161-4170, 2017.
[82] T.-W. Hui, X. Tang, and X. Tang, "LiteFlowNet: A lightweight convolutional neural network for optical flow estimation," *Proceedings of the IEEE conference on computer vision and pattern recognition,* pp. 8981-8989, 2018.
[83] D. Sun, X. Yang, M.-Y. Liu, and J. Kautz, "PWC-Net: CNNs for optical flow using pyramid, warping, and cost Volume," *Proceedings of the IEEE Conference on Computer Vision and Pattern Recognition*, pp. 8934-8943, 2018.
[84] B. Peng, Y. Xian, and Y. Xian, "A convolution neural network-based speckle tracking method for ultrasound elastography," *IEEE International Ultrasonics Symposium,* pp. 206-212, 2018.
[85] M. G. Kibria and H. Rivaz, "GLUENet: Ultrasound Elastography Using Convolutional Neural Network," in *Simulation, Image Processing, and Ultrasound Systems for Assisted Diagnosis and Navigation*, 2018, pp. 21-28.
[86] A. K. Z. Tehrani and H. Rivaz, "Displacement estimation in ultrasound elastography using pyramidal convolutional neural network," *IEEE Transactions on Ultrasonics, Ferroelectrics, and Frequency Control,* p. in press, 2020.
[87] B. Peng, Y. Xian, Q. Zhang, and J. Jiang, "Neural network-based motion tracking for breast ultrasound strain elastography: An initial assessment of performance and feasibility," *Ultrason Imaging,* p. 161734620902527, Jan 30 2020.
[88] S. Wu, Z. Gao, J. Luo, Z. Liu, H. Zhang, and S. Li, "Direct reconstruction of ultrasound elastography using an end-to-end deep neural network," *International Conference on Medical Image Computing and Computer-Assisted Intervention,* pp. 374-382, 2018.
[89] Z. Gao, S. Wu, Z. Liu, J. Luo, H. Zhang, M. Gong, and S. Li, "Learning the implicit strain reconstruction in ultrasound elastography using privileged information," *Med Image Anal,* vol. 58, p. 101534, Dec 2019.
[90] A. Østvik, E. Smistad, T. Espeland, E. A. R. Berg, and L. Lovstakken, "Automatic myocardial strain imaging in echocardiography using deep learning," in *Deep Learning in Medical Image Analysis and Multimodal Learning for Clinical Decision Support*: Springer, 2018, pp. 309-316.
[91] A. Lu, M. Zontak, N. Parajuli, J. C. Stendahl, N. Boutagy, M. Eberle, I. Alkhalil, M. O'Donnell, A. J. Sinusas, and J. S. Duncan, "Learning-based spatiotemporal regularization and integration of tracking methods for regional 4D cardiac deformation analysis," in *International Conference on Medical Image Computing and Computer-Assisted Intervention*, 2017, pp. 323-331.
[92] Q. Zhang, Y. Xiao, W. Dai, J. Suo, C. Wang, J. Shi, and H. Zheng, "Deep learning based classification of breast tumors with shear-wave elastography," *Ultrasonics,* vol. 72, pp. 150-157, Dec 2016.
[93] Q. Zhang, S. Song, Y. Xiao, S. Chen, J. Shi, and H. Zheng, "Dual-mode artificially-intelligent diagnosis of breast tumours in shear-wave elastography and B-mode ultrasound using deep polynomial networks," *Med Eng Phys,* vol. 64, pp. 1-6, Feb 2019.
[94] L. Q. Zhou, J. Y. Wang, S. Y. Yu, G. G. Wu, Q. Wei, Y. B. Deng, X. L. Wu, X. W. Cui, and C. F. Dietrich, "Artificial intelligence in medical imaging of the liver," *World J Gastroenterol,* vol. 25, no. 6, pp. 672-682, Feb 14 2019.

[95] K. Wang, X. Lu, H. Zhou, Y. Gao, J. Zheng, M. Tong, C. Wu, C. Liu, L. Huang, T. Jiang, F. Meng, Y. Lu, H. Ai, X. Y. Xie, L. P. Yin, P. Liang, J. Tian, and R. Zheng, "Deep learning Radiomics of shear wave elastography significantly improved diagnostic performance for assessing liver fibrosis in chronic hepatitis B: a prospective multicentre study," *Gut*, vol. 68, no. 4, pp. 729-741, Apr 2019.

[96] C. Le Berre, W. J. Sandborn, S. Aridhi, M. D. Devignes, L. Fournier, M. Smail-Tabbone, S. Danese, and L. Peyrin-Biroulet, "Application of artificial intelligence to gastroenterology and hepatology," *Gastroenterology,* vol. 158, no. 1, pp. 76-94 e2, Jan 2020.

[97] P. Qin, K. Wu, Y. Hu, J. Zeng, and X. Chai, "Diagnosis of benign and malignant thyroid nodules using combined conventional ultrasound and ultrasound elasticity imaging," *IEEE J Biomed Health Inform*, vol. 24, no. 4, pp. 1028-1036, Apr 2020.

[98] J. Gao, L. Xu, A. Bouakaz, and M. Wan, "A deep siamese-based plantar fasciitis classification method using shear wave elastography," *IEEE Access,* vol. 7, pp. 130999-131007, 2019.

[99] R. R. Wildeboer, C. K. Mannaerts, R. J. G. van Sloun, L. Budaus, D. Tilki, H. Wijkstra, G. Salomon, and M. Mischi, "Automated multiparametric localization of prostate cancer based on B-mode, shear-wave elastography, and contrast-enhanced ultrasound radiomics," *Eur Radiol,* vol. 30, no. 2, pp. 806-815, Feb 2020.

[100] Y. Zhou, J. Xu, Q. Liu, C. Li, Z. Liu, M. Wang, H. Zheng, and S. Wang, "A Radiomics approach with CNN for shear-wave elastography breast tumor classification," *IEEE Trans Biomed Eng,* vol. 65, no. 9, pp. 1935-1942, Sep 2018.

[101] S. H. Yeo, G. R. Kim, S. H. Lee, and W. K. Moon, "Comparison of ultrasound elastography and color Doppler ultrasonography for distinguishing small triple-negative breast cancer from fibroadenoma," *J Ultrasound Med,* vol. 37, no. 9, pp. 2135-2146, Sep 2018.

[102] Y. Xiao, J. Zeng, L. Niu, Q. Zeng, T. Wu, C. Wang, R. Zheng, and H. Zheng, "Computer-aided diagnosis based on quantitative elastographic features with supersonic shear wave imaging," *Ultrasound Med Biol,* vol. 40, no. 2, pp. 275-286, Feb 2014.

[103] R. G. Barr, "Future of breast elastography," *Ultrasonography,* vol. 38, no. 2, pp. 93-105, Apr 2019.

[104] R. G. Barr, "Liver elastography still in its infancy," *Radiology,* vol. 288, no. 1, pp. 107-108, Jul 2018.

[105] Y. N. Zhang, K. J. Fowler, A. Ozturk, C. K. Potu, A. L. Louie, V. Montes, W. C. Henderson, K. Wang, M. P. Andre, A. E. Samir, and C. B. Sirlin, "Liver fibrosis imaging: A clinical review of ultrasound and magnetic resonance elastography," *J Magn Reson Imaging,* vol. 51, no. 1, pp. 25-42, Jan 2020.

[106] A. Tang, G. Cloutier, N. M. Szeverenyi, and C. B. Sirlin, "Ultrasound Elastography and MR Elastography for Assessing Liver Fibrosis: Part 2, Diagnostic Performance, Confounders, and Future Directions," *American Journal of Roentgenology,* vol. 205, no. 1, pp. 33-40, 2015.

[107] A. Treacher, D. Beauchamp, B. Quadri, D. Fetzer, A. Vij, T. Yokoo, and A. Montillo, "Deep learning convolutional neural networks for the estimation of liver fibrosis severity from ultrasound texture," *Medical Imaging 2019: Computer-Aided Diagnosis,* vol. 10950, p. 109503E, Feb 2019.

[108] N. Nishida, M. Yamakawa, T. Shiina, and M. Kudo, "Current status and perspectives for computer-aided ultrasonic diagnosis of liver lesions using deep learning technology," *Hepatol Int,* vol. 13, no. 4, pp. 416-421, Jul 2019.

[109] X. Liu, J. L. Song, S. H. Wang, J. W. Zhao, and Y. Q. Chen, "Learning to diagnose cirrhosis with liver capsule guided ultrasound image classification," *Sensors,* vol. 17, no. 1, p. 149, Jan 13 2017.
[110] D. Meng, L. Zhang, G. Cao, W. Cao, G. Zhang, and B. Hu, "Liver fibrosis classification based on transfer learning and FCNet for ultrasound images," *IEEE Access,* vol. 5, pp. 5804-5810, 2017.
[111] M. Biswas, V. Kuppili, D. R. Edla, H. S. Suri, L. Saba, R. T. Marinhoe, J. M. Sanches, and J. S. Suri, "Symtosis: A liver ultrasound tissue characterization and risk stratification in optimized deep learning paradigm," *Comput Methods Programs Biomed,* vol. 155, pp. 165-177, Mar 2018.
[112] Z. Akkus, J. Cai, A. Boonrod, A. Zeinoddini, A. D. Weston, K. A. Philbrick, and B. J. Erickson, "A survey of deep-learning applications in ultrasound: artificial intelligence-powered ultrasound for improving clinical workflow," *J Am Coll Radiol,* vol. 16, no. 9 Pt B, pp. 1318-1328, Sep 2019.
[113] I. Gatos, S. Tsantis, S. Spiliopoulos, D. Karnabatidis, I. Theotokas, P. Zoumpoulis, T. Loupas, J. D. Hazle, and G. C. Kagadis, "A machine-learning algorithm toward color analysis for chronic liver disease classification, employing ultrasound shear wave elastography," *Ultrasound Med Biol,* vol. 43, no. 9, pp. 1797-1810, Sep 2017.
[114] R. Wei, J. Wang, X. Wang, G. Xie, Y. Wang, H. Zhang, C. Y. Peng, C. Rajani, S. Kwee, P. Liu, and W. Jia, "Clinical prediction of HBV and HCV related hepatic fibrosis using machine learning," *EBioMedicine,* vol. 35, pp. 124-132, Sep 2018.
[115] G. C. Kagadis, P. Drazinos, I. Gatos, S. Tsantis, P. Papadimitroulas, S. Spiliopoulos, D. Karnabatidis, I. Theotokas, P. Zoumpoulis, and J. D. Hazle, "Deep learning networks on chronic liver disease assessment with fine-tuning of shear wave elastography image sequences," *Physics in Medicine & Biology,* in press 2020.
[116] I. Gatos, S. Tsantis, S. Spiliopoulos, D. Karnabatidis, I. Theotokas, P. Zoumpoulis, T. Loupas, J. D. Hazle, and G. C. Kagadis, "Temporal stability assessment in shear wave elasticity images validated by deep learning neural network for chronic liver disease fibrosis stage assessment," *Med Phys,* vol. 46, no. 5, pp. 2298-2309, May 2019.
[117] D. T. Nguyen, T. D. Pham, G. Batchuluun, H. S. Yoon, and K. R. Park, "Artificial intelligence-based thyroid nodule classification using information from spatial and frequency domains," *Journal of clinical medicine,* vol. 8, no. 11, p. 1976, Nov 14 2019.
[118] A. Saftoiu, P. Vilmann, F. Gorunescu, J. Janssen, M. Hocke, M. Larsen, J. Iglesias-Garcia, P. Arcidiacono, U. Will, M. Giovannini, C. F. Dietrich, R. Havre, C. Gheorghe, C. McKay, D. I. Gheonea, T. Ciurea, and E. U. S. E. M. S. G. European, "Efficacy of an artificial neural network-based approach to endoscopic ultrasound elastography in diagnosis of focal pancreatic masses," *Clin Gastroenterol Hepatol,* vol. 10, no. 1, pp. 84-90 e1, Jan 2012.
[119] A. A. Nair, K. N. Washington, T. D. Tran, A. Reiter, and M. A. L. Bell, "Deep learning to obtain simultaneous image and segmentation outputs from a single input of raw ultrasound channel data," *IEEE Trans Ultrason Ferroelectr Freq Control,* in press May 11 2020.
[120] D. Patel, R. Tibrewala, A. Vega, L. Dong, N. Hugenberg, and A. A. Oberai, "Circumventing the solution of inverse problems in mechanics through deep learning: Application to elasticity imaging," *Computer Methods in Applied Mechanics and Engineering,* vol. 353, pp. 448-466, 2019.
[121] G. Wang, "A perspective on deep imaging," *IEEE Access,* vol. 4, pp. 8914-8924, 2016.
[122] R. L. Ling, "A computer generated aid for cluster analysis," *Communications of the ACM,* vol. 16, no. 6, pp. 355-361, 1973.

[123] M. Gasse, F. Millioz, E. Roux, D. Garcia, H. Liebgott, and D. Friboulet, "High-Quality Plane Wave Compounding Using Convolutional Neural Networks," *IEEE Trans Ultrason Ferroelectr Freq Control,* vol. 64, no. 10, pp. 1637-1639, Oct 2017.
[124] N. Awasthi, K. R. Prabhakar, S. K. Kalva, M. Pramanik, R. V. Babu, and P. K. Yalavarthy, "PA-Fuse: deep supervised approach for the fusion of photoacoustic images with distinct reconstruction characteristics," *Biomed Opt Express,* vol. 10, no. 5, pp. 2227-2243, May 1 2019.
[125] S. Gutta, V. S. Kadimesetty, S. K. Kalva, M. Pramanik, S. Ganapathy, and P. K. Yalavarthy, "Deep neural network-based bandwidth enhancement of photoacoustic data," *J Biomed Opt,* vol. 22, no. 11, pp. 1-7, Nov 2017.
[126] Q. Zhang, J. Suo, W. Chang, J. Shi, and M. Chen, "Dual-modal computer-assisted evaluation of axillary lymph node metastasis in breast cancer patients on both real-time elastography and B-mode ultrasound," *Eur J Radiol,* vol. 95, pp. 66-74, Oct 2017.
[127] O. Ronneberger, P. Fischer, and T. Brox, "U-net: Convolutional networks for biomedical image segmentation," in *International Conference on Medical image computing and computer-assisted intervention,* 2015, pp. 234-241.
[128] F. Pesapane, M. Codari, and F. Sardanelli, "Artificial intelligence in medical imaging: threat or opportunity? Radiologists again at the forefront of innovation in medicine," *Eur Radiol Exp,* vol. 2, no. 1, p. 35, Oct 24 2018.